\documentclass{appolb}

\usepackage{graphicx}
\usepackage{subfigure}
\usepackage{cite}
\usepackage{epsfig}
\usepackage{amsmath,amssymb}

\def\tr{{\rm tr}}
\def\vp{{\vec p}}
\def\vx{{\vec x}}

\def\MeV{\,{\rm MeV}}
\def\GeV{\,{\rm GeV}}
\def\fm{\,{\rm fm}}
\def\SU{{\rm SU}}
\def\DeltaA{{\mathcal A}}
\newcommand{\ignore}[1]{}

\begin{document}

\title{Quark Hadron Duality at Finite Temperature 
\thanks{Presented by ERA at the {\em 
LIV Cracow School of Theoretical Physics on QCD meets experiment}, 
June 12-20, 2014.} 
%\thanks{Supported by Spanish DGI grant
%    FIS2011-24149, Junta de Andaluc{\'\i}a grant
%    FQM225, FPA2011-25948 and the JdC Program of the
%    Spanish MICINN.}
} \author{ E. Ruiz Arriola, L.L. Salcedo
  \address{Departamento de F\'{\i}sica At\'omica, Molecular y Nuclear
    \\ and Instituto Carlos I de F\'{\i}sica Te\'orica y
    Computacional, \\ Universidad de Granada, E-18071 Granada, Spain}
  \and E. Meg\'{\i}as$^{\dagger,\ddag}$ \address{$^{\dagger}$Grup de F\'{\i}sica Te\`orica and IFAE,
    Departament de F\'{\i}sica, \\ Universitat Aut\`onoma de Barcelona,
    Bellaterra E-08193 Barcelona, Spain \\
    $^{\ddag}$Max-Planck-Institut f\"ur Physik (Werner-Heisenberg-Institut), F\"ohringer Ring 6, D-80805 Munich, Germany}
}
\maketitle
\begin{abstract}
At low temperatures we expect that all QCD observables are defined in
terms of hadrons. This includes the partition function as well as the
Polyakov loop in all representations.  We analyze the physics
underlying a microscopic derivation of the hadron resonance gas. 
\end{abstract}

\PACS{12.38.Lg, 11.30, 12.38.-t}
 
\bigskip
\bigskip
\section{Introduction}

The promise of a phase transition at finite temperature in QCD from
the confined hadronic phase to a deconfined quark-gluon phase has
triggered a lot of activity both theoretically as well as
experimentally. This original prediction by early lattice
calculations~\cite{Kogut:1982rt,Polonyi:1984zt,Pisarski:1983ms} has
turned into a smooth crossover after many years of accumulated
experience~\cite{Aoki:2006we}. The determination and understanding of the QCD
equation of state (EoS) from the lattice~\cite{Philipsen:2012nu} plays
a crucial role in the current analysis of ultra-relativistic heavy ion
collisions~\cite{Florkowski:2010zz}. 

While in the limit of massless quarks the QCD Lagrangian is scale
invariant, implying a vanishing trace of the energy momentum tensor,
the symmetry is broken explicitly by the finite quark masses and
anomalously by the necessary renormalization which introduces a energy
scale and generates a trace anomaly. The most recent up-to-date
results for 2+1 flavor lattices have been obtained by the
Wuppertal-Budapest(WB)~\cite{Borsanyi:2013bia} and
HotQCD~\cite{Bazavov:2014pvz} collaborations and after continued
discrepancies there is a final consensus that maximal violation of
scale invariance occurs at $T_s \sim 200\MeV$ where the trace anomaly
reaches its maximum value.

At very high temperatures, the typical momentum scales or $\mu \sim 2 \pi T$
are large and finite quark mass effects can be neglected.  Due to asymptotic
freedom the strong and running coupling constant becomes small and thus
interactions can be neglected. Thus, one effectively has a gas of free and
massless $4 N_f N_c $ fermions (quarks and anti-quarks) and $2 (N_c^2-1)$
bosons (gluons), and scale invariance is restored. At the same time colour is
delocalized corresponding to a deconfined phase.  This allows in principle to
count the number of $2 N_f N_c $ quark and $2 (N_c^2-1)$ gluon elementary
species by means of the Stefan-Boltzmann law.  Current analyses show that this
happens for temperatures much larger than $T_s$.

Yet, there is the firm belief that because of confinement, hadron
states, composite, extended and most often unstable bound states made
of quarks and gluons build a complete basis at very low temperatures,
where they effectively behave as point-like, stable, non-interacting
and structureless particles. This quark-hadron duality resembles to a
large extent the similar duality between different degrees of freedom
as the one found in deep inelastic scattering (for a review see
\cite{Melnitchouk:2005zr}). It is remarkable that in the large $N_c$
limit some of these requirements are indeed fulfilled (for a review
see e.g.~\cite{Lucini:2012gg}), except and most notably the point-like
character. Of course, as the temperature is raised we expect many
excited states to contribute, but also that finite width and finite
size effects play a role.  How this hadronization happens in detail
has not really been understood so far. Lattice calculations suggest
that there is a smooth transition or crossover from the purely
hadronic phase to the purely quark-gluon phase~\cite{Aoki:2006we}.
Actually, it is not obvious when this hadronic picture fails in
practice or what is the main mechanism behind quark and gluon
liberation at the lowest possible temperature. The present
contribution tries to address this problem guided by our own
experience on the field.

\section{The hadron resonance gas}

In the opposite limit of very low temperatures $T \ll T_s$
one expects an interacting gas of the lightest colour neutral
particles (for $N_f=3$ pions and kaons)~\cite{Gerber:1988tt}. Because
of spontaneous chiral symmetry breaking, the low-lying pseudoscalar
particles are the lightest Goldstone bosons made of $u,d,s$ quarks. For
the typical low momenta in the heat bath $p \sim 2 \pi T \ll
m_\pi,m_K$ Goldstone bosons interact weakly through derivative
couplings and hence interactions are strongly suppressed.  Thus scale
invariance violations of the non-interacting gas are due to the finite
pion and kaon masses.  Therefore, in the chiral limit of massless
pseudoscalars scale invariance is also exact at sufficiently low
temperatures. Because of the small signal, current lattice
calculations of the trace anomaly are just above the edge of this pion
and kaon gas, which is expected to work for $ T \sim m_\pi /2\pi $.

\begin{figure}[h]
\epsfig{figure=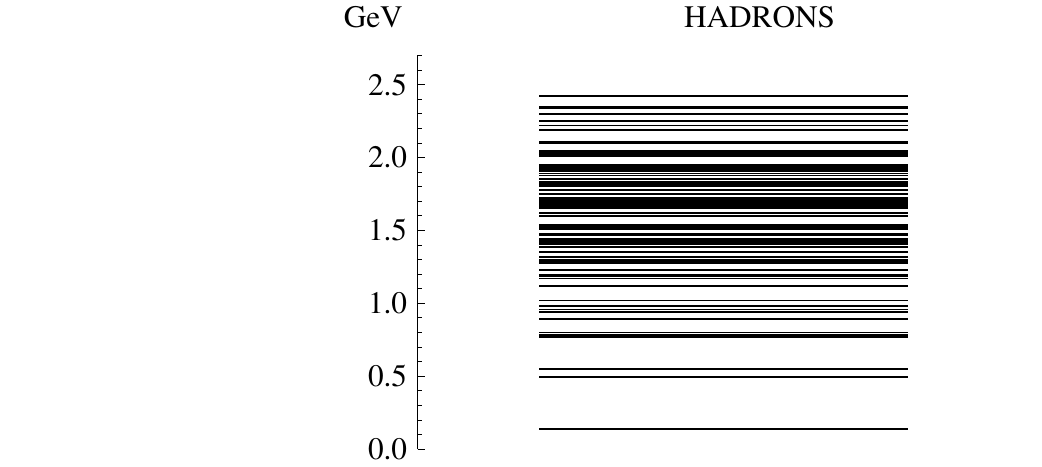,height=5cm,width=5cm,angle=0}
%\hskip1cm 
\epsfig{figure=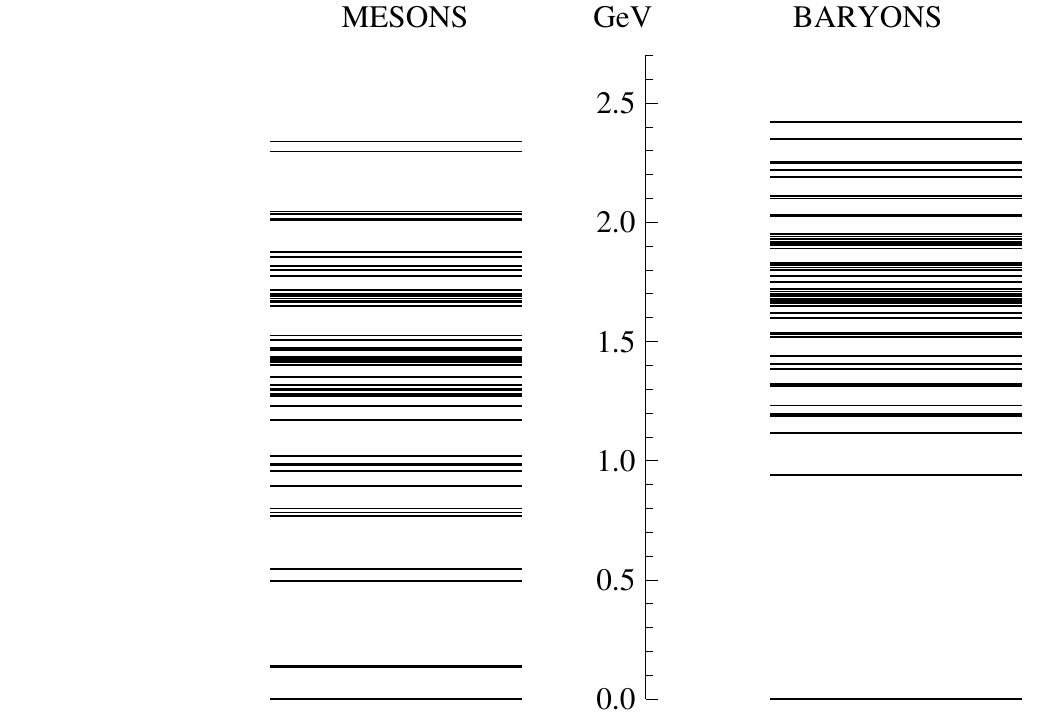,height=5cm,width=6cm,angle=0} 
\caption{Left: The full hadron spectrum made of $u,d,s$ quarks from the
  PDG~\cite{Nakamura:2010zzi}. Right: Mesonic and Baryonic spectrum.
\label{fig:spectrum}}
\end{figure}

When the temperature is raised, hadronic interactions among pions
start playing a role and two- and more particle states contribute to
the thermodynamic properties. The calculation may be organized
according to the quantum virial expansion~\cite{Dashen:1969ep} where
there are two kinds of contributions. The excluded volume corrections
come from repulsive interactions which prevent particles to approach
each other below a certain distance. The resonance contributions stem
from attractive interactions which generate states living long enough
to produce pressure in the system, meaning that the resonance can hit
the wall of the container before it decays. A well known example is
$\pi\pi$ scattering where one has attractive and resonating states in
the isospin $I=0,1$ corresponding to the $\sigma$ and $\rho$
resonances whereas one has a repulsive core in the $I=2$ exotic
channel~\cite{Venugopalan:1992hy,Kostyuk:2000nx} providing a measure
of the finite pion size. Once a $\rho$ meson is created, it may
interact with another pion and produce a resonance, $\rho \pi \to A_1$
(which is a $3\pi$ state) and so on. For baryons, the situation is
similar where $N \pi \to \Delta$ is a good example of a resonance
contribution. This separation between attractive and repulsive
contributions leaves out the residual interaction stemming from the
background scattering. The Hadron Resonance Gas (HRG) corresponds to
assume that all interactions among the lightest and stable particles
can be described by an ideal gas of non-interacting resonances which
are effectively pictured as stable, point-like and elementary
particles, hence the trace anomaly is given by
\begin{eqnarray}
\DeltaA_{\rm HRG} (T) \equiv \frac{\epsilon - 3 P}{T^4} = \frac{1}{T^4}\sum_n \int \frac{d^3 p}{(2\pi)^3} \frac{E_n(p)- \vec p \cdot \nabla_p E_n (p)}{e^{E_n(p)/T}+\eta_n} \, , 
\label{eq:tran-HRG}
\end{eqnarray}
where the sum is over {\it all} hadronic states including spin-isospin
and anti-particle degeneracies, $\eta_n=\mp 1$ for mesons and baryons
respectively, $E_n(p)= \sqrt{p^2+M_n^2}$ is the energy and $M_n$ are
the hadron masses. This collection of masses is usually called the
{\it hadron spectrum} and most often identified with the PDG
compilation which represents an established consensus among particle
physicists~\cite{Nakamura:2010zzi}. The states classification echoes
the non-relativistic quark model, namely mesons are $q \bar q$ states
and baryons $qqq$ states. Those falling outside this category are
classified as ``further states''.  The hadron spectrum obtained from
the PDG is depicted in Fig.~\ref{fig:spectrum} as well as its
separation into mesonic and baryonic spectra. 

\begin{figure}[h]
\begin{center}
\epsfig{figure= 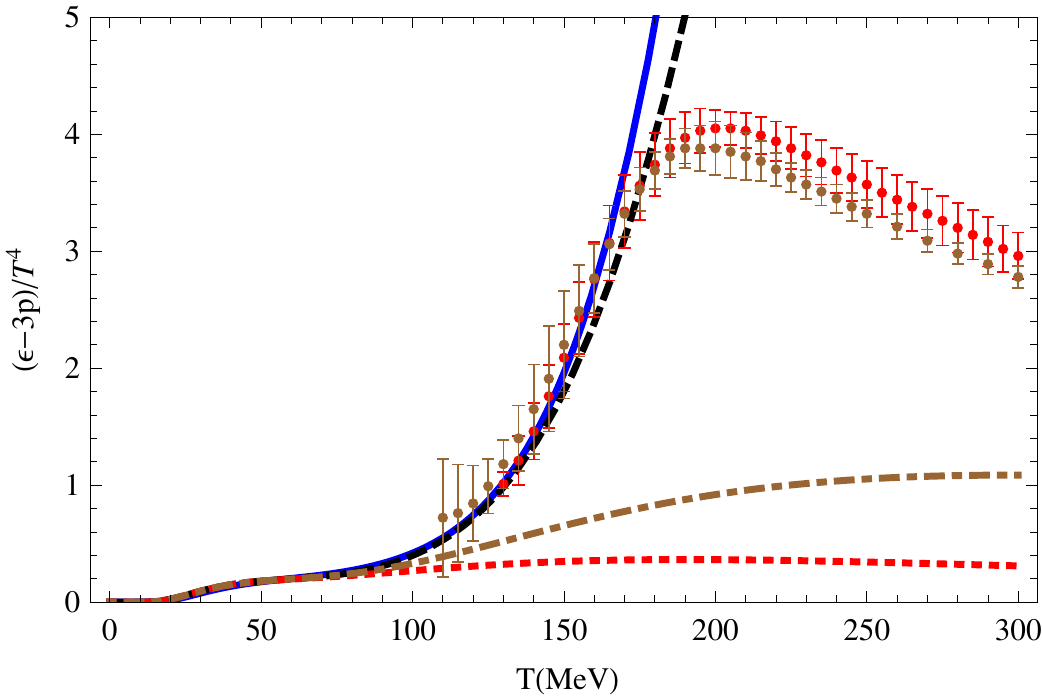,height=6cm,width=6cm,angle=0}
\end{center}
\caption{Trace anomaly for the HRG using the PDG~\cite{Nakamura:2010zzi}
  (dashed) and the RQM~\cite{Godfrey:1985xj,Capstick:1986bm} (full) compared
  to the continuum extrapolated results of the WB~\cite{Borsanyi:2013bia} and
  HotQCD~\cite{Bazavov:2014pvz} collaborations. We also show the contributions
  from states with $M \le 600 \MeV$ (dotted) and with $M \le 800 \MeV$
  (dot-dashed).}
\label{fig:trace-PDG}
\end{figure}

The result of using Eq.~(\ref{eq:tran-HRG}) with the PDG
states~\cite{Nakamura:2010zzi} compared to the continuum extrapolated
results of the Wuppertal-Budapest (WB)~\cite{Borsanyi:2013bia} and
HotQCD~\cite{Bazavov:2014pvz} collaborations is shown in
Fig.~\ref{fig:trace-PDG}. It is amazing that this exceedingly simple
picture works {\it accurately} almost below $T_s$ (the maximal
scale-violating temperature) at about $T \lesssim 170 {\rm MeV}$. Note
that the lowest lattice data points at $T=120 \MeV$ are first
saturated when states with masses above the vector mesons are
included. Higher temperatures start feeling excited hadronic states,
which by their nature embody relativistic dynamics. The quantum
statistical bosonic and baryonic character of the states accounts by
less than $1\%$ correction to the classical Maxwell-Boltzmann
distribution (corresponding to take $\eta_n \to 0$) at about $T=200
{\rm MeV}$, already well beyond the range where lattice and HRG agree
and much smaller than the lattice $10\%$ uncertainties. The transition
from hadrons to quarks has been analyzed on the light of
strangeness~\cite{Bazavov:2013dta} and an observable which vanishes
for the HRG has been proposed.

Actually, the HRG model has arbitrated the discrepancies between the
different collaborations in the past and the lattice community have
had a long struggle until agreement {\it between them} and {\it with
  the HRG model} has been declared.  While this has made the HRG model
a sort of holy grail (see e.g.~\cite{Huovinen:2009yb} and references
therein), it is good to remind that despite of the phenomenological
success it is not a theorem, nor a well defined approximation from the
original QCD Lagrangian. The most compelling argument is that it is to
date unclear how corrections to this simple approach should be
implemented nor what the error estimate for the HRG should be. For a
comparison, we also show the result of Eq.~(\ref{eq:tran-HRG}) using
instead the Relativized Quark Model
(RQM)~\cite{Godfrey:1985xj,Capstick:1986bm} which essentially combines
two basic elements, the static energy among the constituents and a
relativistic form of the kinetic energy which does not consider the
spin of the particles but does not contain more parameters than QCD
itself. The likewise impressive agreement of the RQM trace anomaly
with the lattice data not only illustrates our point on the lack of
uniqueness of what is actually being checked by these comparisons, but
also that the RQM may provide information not listed in the PDG
booklet.

For instance, the PDG lists the quantum numbers, decay modes, masses
and widths of the resonances building the hadron spectrum, but no
information on their size is provided. While for unstable particles
this is a problematic issue (see e.g.~\cite{Sekihara:2012xp} and
references therein), within the heavy-ion literature the assumption of
a constant volume is frequently made (see e.g. \cite{Begun:2012rf}).
This information can be accessed by means of quark models or lattice
calculations.  On the other hand, as discussed in
Ref.~\cite{Arriola:2012vk}, the purely resonance character makes the
very definition of the mass ambiguous, and this allows to generate an
error band on the PDG prediction of the HRG model which gives a spread
for the trace anomaly about half the lattice uncertainties.

\section{QCD at finite temperature}

As a guideline, let us provide some of the main features of QCD at
finite temperature emphasizing some relevant aspects for the
discussion. Many gaps in this sketchy presentation may be filled by
consulting e.g.~\cite{Philipsen:2012nu} and references therein.  The
QCD Lagrangian (in Euclidean spacetime) is given by
\begin{eqnarray}
{\cal L}_{\rm QCD} = -\frac{1}{4} G_{\mu\nu}^a G_{\mu\nu}^a + \sum_f \overline{q}_f^a (i \gamma_{\mu} D_{\mu} - m_f) q_f^a \,.
\end{eqnarray}
The QCD Lagrangian is invariant under colour gauge transformations 
\begin{eqnarray}
&& q (x) \to e^{i \sum_a (\lambda_a)^c \alpha_a (x)} q (x) \equiv \omega(x)
  q(x) \,, \nonumber \\
&& A_\mu(x) \to \omega^{-1}(x) A_\mu(x) \omega(x) +
  \frac{i}{g}\omega^{-1}(x)\partial_\mu \omega(x) \,. 
\nonumber 
\end{eqnarray}
The QCD thermodynamics is obtained from the partition function 
\begin{eqnarray}
Z_{\rm QCD}&=& {\rm Tr} \, e^{-H/T} = \sum_n e^{-E_n/T} \nonumber \\
&=& \int {\cal D} A_{\mu ,a} \exp \left[- \frac1{4} \int d^4 x
(G_{\mu\nu}^a)^2 \right] {\rm Det} (i \gamma_{\mu} D_{\mu} - m_f) \,, \nonumber 
\end{eqnarray}
with the periodic or anti-periodic boundary conditions 
for gluons and quarks respectively 
\begin{eqnarray}
&& q (\vec x, \beta) = - q( \vec x,0)\,, \qquad A_\mu (\vec x , \beta) = A_\mu
(\vec x , 0)\,, \qquad \beta = 1/T \,, \nonumber \\ 
&& \int \frac{dp_0}{2\pi} f(p_0) \to  T \sum_n f (w_n)\,, \nonumber 
%&& w_n = (2n+1 )\pi T\,,  \qquad w_n = 2 n \pi T \,.  \nonumber  
\end{eqnarray}
where the Matsubara frequencies are $ w_n = 2 n \pi T$ for gluons and
$w_n = (2n+1) \pi T$ for quarks.
%This corresponds to taking the Matsubara frequencies. 
Preservation of the quark antiperiodic boundary conditions implies
that only periodic gauge transformations are allowed, namely
\begin{eqnarray}
\omega({\vec x},x_0+\beta) = \omega({\vec x}, x_0) \,, \qquad \beta=1/T \,. 
\end{eqnarray}
Within the convenient Polyakov gauge ($A_0(x)$ stationary and everywhere
diagonal \cite{Suzuki:1994ay}) the most general remaining transformation is
either stationary and diagonal or of the type
\begin{eqnarray}
\omega(x_0)= e^{i2\pi x_0 \lambda /\beta}\,, 
\end{eqnarray}
where $\lambda = {\rm diag}(n_1,\cdots,n_{N_c}) $ and ${\rm Tr} \,
\lambda =0 $.  Large Gauge Invariance implies periodicity in $A_0$
with period $2\pi T/g$
\begin{eqnarray}
A_0 \to A_0 -\frac{ 2\pi T}{g} \lambda
\,.
\end{eqnarray}
This is the finite temperature version of the Gribov copies, i.e., the fact
that there is no complete gauge fixing in a non-abelian gauge theory.  A
drastic consequence of this periodicity property is that it becomes explicitly
Broken in perturbation theory \cite{Dunne:1996yb,Salcedo:1998sv}. Thus, we may
consider this as a signal of the relevance of non-perturbative finite
temperature gluons.

In the limit of massless quarks ($m_f = 0$) the QCD Lagrangian is scale
invariant, i.e.  $ x \longrightarrow \mathbf{\lambda} x$.  This symmetry is
broken by quantum corrections due to the necessary regularization.  To see
this consider the partition function dependence on the coupling constant $g$
after the rescaling of the gluon field $ \bar A_\mu = g A_\mu$ (and ignoring
renormalization issues)
\begin{eqnarray}
Z=\int {\cal D} \bar A_{\mu ,a} \exp \left[- \frac1{4 g^2} \int d^4 x
(\bar G_{\mu\nu}^a)^2 \right] {\rm Det} (i \gamma_{\mu} D_{\mu} )
.
\end{eqnarray}
Note that the {\it only} dependence on $g$ is the one shown explicitly. 
Thus,
\begin{eqnarray}
\frac{\partial \log Z}{\partial g } = \frac{1}{2g^3}
\left\langle \int d^4 x
(\bar{G}_{\mu\nu}^a)^2 \right\rangle 
= \frac{1}{2g}\frac{V}{T}\langle  (G_{\mu\nu}^a)^2
\rangle_T 
\end{eqnarray}
where in the last equation we have assumed a vacuum space time independent
configuration, with $V$ the volume of the system. On the other hand
the free energy and internal energy are given by the thermodynamic
relations
\begin{eqnarray}
  F = -PV = - T \log Z  \,, \qquad  
  \epsilon = \frac{E}{V} =  \frac{T^2}{V} 
\frac{\partial  \log Z }{\partial T} \,,
\end{eqnarray}
and the trace anomaly becomes   
\begin{eqnarray}
\epsilon- 3 P  = T^5 \frac{\partial}{\partial T}
\left( \frac{P}{T^4} \right) \,. 
\label{eq:em3p}
\end{eqnarray}
Generally, a renormalization scale $\mu$ has to be introduced to handle both 
IR and UV divergences. Thus, on purely dimensional grounds one has 
\begin{eqnarray}
\frac{P}{T^4}= f(g(\mu),\log(\mu/2\pi T)) \,.
\end{eqnarray}
Physical results should not depend on the renormalization scale, thus
using that $ d P / d \mu =0$ we get   
\begin{eqnarray}
\frac{\partial}{\partial \log T}\left(\frac{P}{T^4}\right) =
\frac{\partial g}{\partial\log\mu}\frac{\partial}{\partial g}
\left(\frac{P}{T^4} \right) \, .  
\end{eqnarray}
Introducing the beta function
\begin{eqnarray}
\beta (g) = \mu \frac{d g}{d \mu } = - \beta_0 g^3 +  {\cal O}(g^5) \,,
\qquad
\beta_0 =   \frac{11 N_c-2N_f}{48 \pi^2} \,,
\end{eqnarray}
the trace anomaly becomes then
\begin{eqnarray}
\epsilon - 3 P = \langle G^2\rangle_T - \langle G^2\rangle_0
\qquad \text{(massless quarks)}
\label{eq:G2}
\end{eqnarray}
where
\begin{eqnarray}
G^2 =  \frac{\beta(g)}{ 2 g } (G_{\mu\nu}^a)^2 
\,.
\end{eqnarray}
Here we have implemented, in full harmony with standard lattice practice (see,
e.g., \cite{Boyd:1996bx}), a subtraction to renormalize the vacuum
contribution \cite{Leutwyler:1992cd,Miller:2006hr}. The vanishing of the trace
anomaly at zero temperature is consistent with quark-hadron duality, see Eq.~(\ref{eq:tran-HRG}).

Also in the massless quark limit, the QCD Lagrangian 
is invariant under $SU_R (N_f) \times SU_L (N_f)  $ chiral
transformations,
\begin{eqnarray}
 q (x) \to e^{i \sum_a (\lambda_a)^f \alpha_a} q (x) \,,\qquad 
q (x) \to e^{i \sum_a (\lambda_a)^f \alpha_a\gamma_5} q (x) 
\,.
\end{eqnarray}
Chiral symmetry is spontaneously broken by the chiral condensate in
the vacuum down to $SU_V (N_f)$
\begin{eqnarray}
\langle \bar q q  \rangle \neq 0 \,.
\end{eqnarray}
The Gell-Mann--Oakes--Renner relation, which in the simplest $N_f=2$
case becomes  
\begin{eqnarray}
2 \langle \bar q q \rangle m_q = -f_\pi^2 m_\pi^2
\,,
\end{eqnarray}
relates the quark condensate and the quark mass to physically measurable
quantities such as the pion mass $m_\pi$ and the pion weak decay constant
$f_\pi$. This relation exemplifies the {\it quark-hadron duality}, namely the
fact that, in the confined phase of QCD, we expect quark observables to be
representable by hadronic observables.

In the opposite limit that all quarks become infinitely massive, $m_f \to
\infty$, the quark determinant becomes a constant which can be
factored out of the path integral
\begin{eqnarray}
Z_{\rm QCD} \to Z_{\rm YM} {\rm Det} (- m_f)  \,,
\end{eqnarray}
and the resulting action corresponds to a pure Yang-Mills theory 
\begin{eqnarray}
Z_{\rm YM}  = \int {\cal D} A_{\mu ,a} \exp \left[- \frac1{4} \int d^4 x
(G_{\mu\nu}^a)^2 \right]  
\,.
\end{eqnarray}
This purely gluonic action exhibits a larger symmetry: 't Hooft Center
Symmetry ${\mathbb Z}(N_c)$, namely, invariance under gauge transformations
which are periodic modulo a center element of $\SU(N_c)$
\begin{eqnarray}
\omega(\vec{x},x_0+\beta) = z \, \omega(\vec{x},x_0) \,, \qquad z^{N_c}=1 ,  \quad 
(z\in {\mathbb Z}(N_c))\,.
\end{eqnarray}
An example for $z=e^{i2\pi k/N_c}$ and in the Polyakov gauge is given by the choice
\begin{eqnarray}
 \omega(x_0) = e^{i2\pi x_0 k \lambda/(N_c\beta)} \,, \qquad
A_0 \rightarrow A_0 - \frac{2\pi T}{gN_c} k \lambda \,,
\end{eqnarray}
with $\lambda={\rm diag}(1-N_c,1,\ldots,1)$.  The Polyakov loop is an
order parameter of the center symmetry which is related to the free
energy of a colour charge in the medium. In the Polyakov gauge, the
Polyakov loop in the fundamental representation reads
\begin{eqnarray}
\Omega_F (\vec x) = e^{i g A_0 (\vec x)/T}
\,, \qquad A_0 = \sum_{a=1}^{N_c^2-1} \lambda_a A_0^a 
\,.
\end{eqnarray}
The vacuum expectation value of the Polyakov loop transforms under the
previous gauge transformation as
\begin{eqnarray}
L_T = \langle {\rm tr}_c e^{i g A_0/T} \rangle = e^{-F_q /T}  ~\longrightarrow~
z L_T \,.
\end{eqnarray}
Therefore unbroken symmetry ($L_T=zL_T$)) implies $L_T=0$ and hence
$F_q = \infty $. The divergence of the free energy of a colour charge
in the fundamental representation is interpreted as a signal for
confinement \cite{Svetitsky:1985ye}. The renormalization of the
Polyakov loop is a subtle issue addressed in the lattice in
Ref.~\cite{Kaczmarek:2002mc}.

In gluodynamics the center symmetry is spontaneously broken above a
critical temperature and $L_T/N_c$ approaches unity (or any other
central element) as the temperature increases. In fact, at high
temperatures $A_0 \ll T$ and one may expand~\cite{Megias:2005ve},
\begin{eqnarray}
\frac{1}{N_c}L_T = 1 - \frac{g^2\langle {\rm tr}_c A_0^2 \rangle}{ 2 N_c T^2} 
+ \dots = 
e^{-\frac{g^2\langle {\rm tr}_c A_0^2 \rangle}{ 2 N_c T^2} + \cdots}  
\label{eq:LT-dim2}
\end{eqnarray}
Note that while this formula suggests that $L_T \le N_c $, the
renormalization overshoots this value at a perturbative level in a
tiny but visible way~\cite{Gava:1981qd,Kaczmarek:2005uv,Megias:2005ve}.

In full QCD the center symmetry is explicitly broken, which results in $L_T >
0$ at all temperatures. In the limit of heavy quarks and low temperature one
has $L_T = {\cal O} (e^{-m_q/T}) \ll 1 $ or $F_q \to m_q + \epsilon + \dots $,
where $\epsilon$ is the smallest residual binding energy of doubly heavy $Q
\bar Q$ meson. Likewise for light dynamical quarks $L_T \sim e^{-\Delta /T}$
where $\Delta$ denotes the ground state mass of a heavy-light $Q \bar q$ meson
(the mass of the heavy quark excluded)~\cite{Megias:2012kb}.

One can also define Polyakov loops in higher representations which
have been subject of attention only a few times despite its very
interesting
properties~\cite{Gupta:2007ax,Mykkanen:2012ri,Megias:2013xaa,Megias:2014bfa}.

Scale invariance is also broken in gluodynamics. For instance in 
perturbation theory to two loops one has \cite{Kapusta:1979fh},
\begin{eqnarray}
\DeltaA \equiv \frac{\epsilon-3P}{T^4} = \frac{N_c(N_c^2-1)}{72}
\beta_0 g(T)^4 + {\cal O}(g^5) \,,
\end{eqnarray}
where at lowest order 
\begin{eqnarray}
\frac{1}{g(\mu)^2} = \beta_0 \log (\mu^2/\Lambda_{\rm QCD}^2)
\,.
\end{eqnarray}
Thus, taking $\mu = 2 \pi T$, we expect to have a free gas of gluons
and quarks in the high temperature limit. In the simple case of
non-interacting particles the partition function is given by
\begin{eqnarray}
\log Z = V \eta g_i \int \frac{d^3 p }{(2\pi)^3} \log \left[ 1+ \eta
  e^{-E_p/T} \right]
\,,
 \qquad E_p = \sqrt{p^2+m^2} \,,
\end{eqnarray}
where $\eta=-1 $ for bosons,  $\eta=+1 $ for fermions and $g_i$ is the number
of species. From the partition function we have the thermodynamic identities
\begin{eqnarray}
 F &=& - T \log Z \,,\qquad 
 P  = - \frac{\partial F}{\partial V}\,, \nonumber \\
 S &=&  - \frac{\partial F}{\partial T}\,, \qquad 
 E  = F + T S \,.
\nonumber
\end{eqnarray}
For the massless quark and gluon gas the pressure is given by 
\begin{eqnarray}
P = \left[ 2 (N_c^2-1) + 4 N_c N_f \frac{7}{8} \right] \frac{\pi^2}{90} T^4 \,,
\end{eqnarray}
which is the Stefan-Boltzmann law. Because in the high temperature limit 
the particles are effectively massless, scale invariance is restored and
hence the (reduced) trace anomaly vanishes
\begin{eqnarray}
\DeltaA \equiv \frac{\epsilon- 3 P}{T^4} \to 0 \qquad ( T \to \infty)   
\,.
\end{eqnarray}
These expectations have been checked in Ref.~\cite{Borsanyi:2012ve} by
a lattice study in a wide temperature window, $T=0.7 \dots 10^3 T_c$.

Thus, the quark condensate $\langle \bar q q \rangle $ and the
Polyakov loop in the fundamental representation $L_T$ become true
order parameters in quite opposite situations. While $\langle \bar q q
\rangle $ signals spontaneous chiral symmetry breaking in the massless
quark limit, $L_T$ signals confinement for infinitely heavy
quarks. The real situation is somewhat intertwined, and can be
summarized as follows.

\begin{itemize} 
\item Order parameter of chiral symmetry breaking ($m_q=0$)  \\
Quark condensate $SU_R (N_f) \times SU_L (N_f) \to SU_V(N_f) $
\[
\langle \bar q q \rangle \neq 0 \quad (T < T_c) \,, \qquad 
\langle \bar q q \rangle = 0 \quad (T > T_c )
\,.
\] 
\item Order parameter of deconfinement   ($m_q=\infty$)  \\
Polyakov loop:  Center symmetry ${\mathbb Z}(N_c)$ broken 
\[
L_T=  0 \quad (T < T_c) \,,
\qquad 
L_T  > 0 \quad (T > T_c )
\,.
\] 
\item In the real world $m_q$ is finite but inflection points nearly coincide 
(accidental?) 
\[
\frac{d^2}{dT^2} L_T=0 
\,,    
\qquad  \frac{d^2}{dT^2} \langle \bar q q \rangle_T =0 
\,,
\]
at about the same temperature $T_c=155(10)\MeV$.
\end{itemize}
The chiral-deconfinement crossover is a unique prediction of lattice
QCD. Whether or not this result is accidental, could be answered by
computing the (connected) crossed correlator,
\begin{eqnarray}
\langle \bar q q \, {\rm tr}_c \, e^{ i g A_0/T}  \rangle 
- \langle \bar q q \rangle \, \langle   {\rm tr}_c \, e^{ i g A_0/T}
   \rangle 
= \frac{\partial L_T}{ \partial m_q} \,, 
\end{eqnarray}
which corresponds to the quark mass dependence of the Polyakov loop.

Finally, the correlation function between Polyakov loops in an
arbitrary representation $R$ exhibits Casimir scaling (quenched approximation)
\cite{Greensite:2003bk}
\begin{eqnarray}
\langle {\rm Tr}_R\Omega (\vec x_1) {\rm Tr}_R \Omega (\vec x_2)^\dagger
\rangle \approx e^{- \sigma_R | \vec x_1 - \vec x_2 |/T}
\,, \qquad
\sigma_R = (C_R/C_F) \sigma_F 
\,.
\label{eq:pol-corr}
\end{eqnarray}
It can be shown that this correlation function, which for the
fundamental representation is related to the singlet $\bar q q$ free
energy, $F_1(r,T)$, can also be written as a ratio of partition
functions between $\bar Q Q$ sources placed at a distance $|\vec
x_1-\vec x_2|$ and the vacuum~\cite{Luscher:2002qv}, hence satisfying
a spectral decomposition with integral weights $w_n$ and positive
energies $E_n(|\vec x_1 - \vec x_2|)>0 $,
\begin{eqnarray}
\langle {\rm Tr}_F\Omega (\vec x_1) {\rm Tr}_F \Omega (\vec x_2)^\dagger
\rangle =\sum_n w_n e^{-E_n(|\vec x_1 - \vec x_2|)/T} = e^{-F_1 (r,T)/T}
\,, 
\label{eq:pol-corr-unq}
\end{eqnarray}
where the singlet free energy $F_1(r,T)$ has been introduced. One
important property is that at large distances (for the unquenched full
QCD case),
\begin{eqnarray}
\langle {\rm Tr}_F\Omega (\vec r) {\rm Tr}_F \Omega
(0)^\dagger \rangle \to  
|\langle {\rm Tr}_F\Omega \rangle |^2 = L_T^2 \,. 
\label{eq:pol-corr-loop}
\end{eqnarray}

\section{Relativized Quark-Gluon models}

\subsection{Relativity and thermodynamics}

One of the most troublesome aspects of hadron binding is that it makes
relatively heavy particles from massless ones, hence most of the mass
comes from the interaction. A prominent example is the Glueball in
gluodynamics, where the lightest $0^{++}$ state~\cite{West:1995ym} has
a mass $M_{0^{++}} /\sqrt{\sigma} \sim 4.5$ while gluons are
massless. Fully relativistic few body equations are not only hard to
handle but encounter many difficulties regarding cluster decomposition
properties~\cite{Coester:1965zz,Keister:1991sb,Keister:2011ie}. This
feature is particularly interesting as it is related to the
compositeness nature of relativistic particles which build the
hadrons, and strictly speaking we have to deal with relativistic
statistical mechanics of interacting particles, a subject which has a
long history~\cite{hakim2011introduction}.

Unfortunately, as we have shown, the physics of finite temperature QCD
below the phase transition involves the excited hadronic spectrum, and
thus relativity becomes an essential ingredient in the
game. Relativized quark models (RQM) combine two basic elements, the
static energy among the constituents and a relativistic form of the
kinetic energy which does not consider the spin of the
particles~\cite{Godfrey:1985xj,Capstick:1986bm}.

\subsection{The linear potential}

In the Born-Oppenheimer approximation the object to be analyzed is the
interaction between heavy sources $A$ and $B$. In perturbation theory
one has one gluon exchange which yields a Coulomb like interaction, 
\begin{eqnarray}
V_{AB}(r) = \lambda_A \cdot \lambda_B \frac{\alpha_s}{r} \,,  
\end{eqnarray}
where $\alpha_s=g^2/(4\pi)$ and $\lambda_A$ and $\lambda_B$ are the
generators\footnote{These are more customarily denoted by $T$, while
  $\lambda$ is twice the generator.} of the SU(3) colour group
corresponding to the representation of the source.  This form of the
colour interaction exhibits Casimir scaling, a property that is
violated only at three loops in perturbation
theory~\cite{Anzai:2010td} and appears to hold non-perturbatively on
the lattice with an additional linear potential
contribution~\cite{Bali:2000un}. Thus, to a good approximation, the
interaction between heavy sources on the lattice reads,
\begin{eqnarray}
V_{AB}(r) = \lambda_A \cdot \lambda_B \left[ \frac{\alpha_s}{r} -
  \kappa r \right]
\,.
\end{eqnarray}
Thus, for quark-antiquark or gluon-gluon pairs coupled to a singlet
state or a quark-quark pair coupled to the antifundamental representation
(diquark) the following relations are obtained
\begin{eqnarray}
V_{Q \bar Q} (r) &=& \sigma_F \, r - \frac{4 \alpha_s}{3r} + \cdots \,, \\
V_{GG} (r) &=& \sigma_A \,  r - \frac{3 \alpha_s}{r} + \cdots  \,, \\ 
V_{QQ} (r) &=& \sigma_d \,  r - \frac{2\alpha_s}{3 r} + \cdots  \,.
\end{eqnarray}
In what follows we use $\sigma$ to denote the string tension $\sigma_F$.
As a consequence of Casimir scaling the ratio between the fundamental $Q
\bar Q \equiv ({\bf 3} \times \bar {\bf 3})_{\bf 1}$, adjoint $GG \equiv ({\bf
  8} \times {\bf 8})_{\bf 1}$ and diquark $QQ\equiv ({\bf 3} \times 
{\bf 3})_{\bar{\bf 3}}$ colour sources are
\begin{eqnarray}
\frac{\sigma_A}{\sigma_F}= \frac{9}{4} \,,
 \qquad \frac{\sigma_d}{\sigma_F}= \frac{1}{2} 
\,.
\end{eqnarray}

By making simplifying assumptions, easy relations can be derived from Casimir
scaling. For instance, consider the lowest glueball state by analyzing two
massless spin-1 particles in the CM system assuming spin independent
interactions, and similarly for the $\rho$ meson as composed of two massless
quarks.  Neglecting the Coulomb term, the respective mass operators appearing
in the Salpeter equation read
\begin{eqnarray}
\hat M_G = 2p + \sigma_A \, r 
\,,
\qquad
\hat M_M = 2p + \sigma_F \, r  \,.
\end{eqnarray}
Simple dimensional considerations imply that the eigenmasses are proportional
to the square root of the string tension, thus
\[
M_{g,0^{++}} /m_\rho \approx 3/2
\,.
\]
Here, as it is customary, we have matched the scales of gluodynamics
and QCD by assuming a common value of $\sigma_F$ in both theories. A
rough estimate of the mass follows from using the uncertainty
principle for the ground state, namely, by taking $p r \sim 1$. For
the glueball this yields
$$
M_0 \approx \min \left[ \frac{2}{r} + \sigma_A \, r \right] 
= 2 \sqrt{2\sigma_A} = 4.2\sqrt{\sigma}
. 
$$

\subsection{The cumulative number}

\begin{figure}[h]
\epsfig{figure=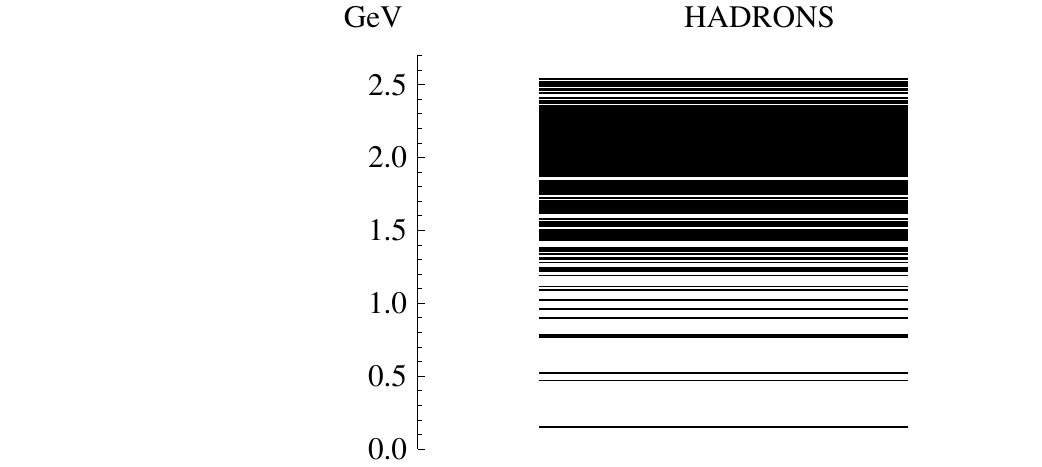,height=5cm,width=5cm,angle=0}
\epsfig{figure=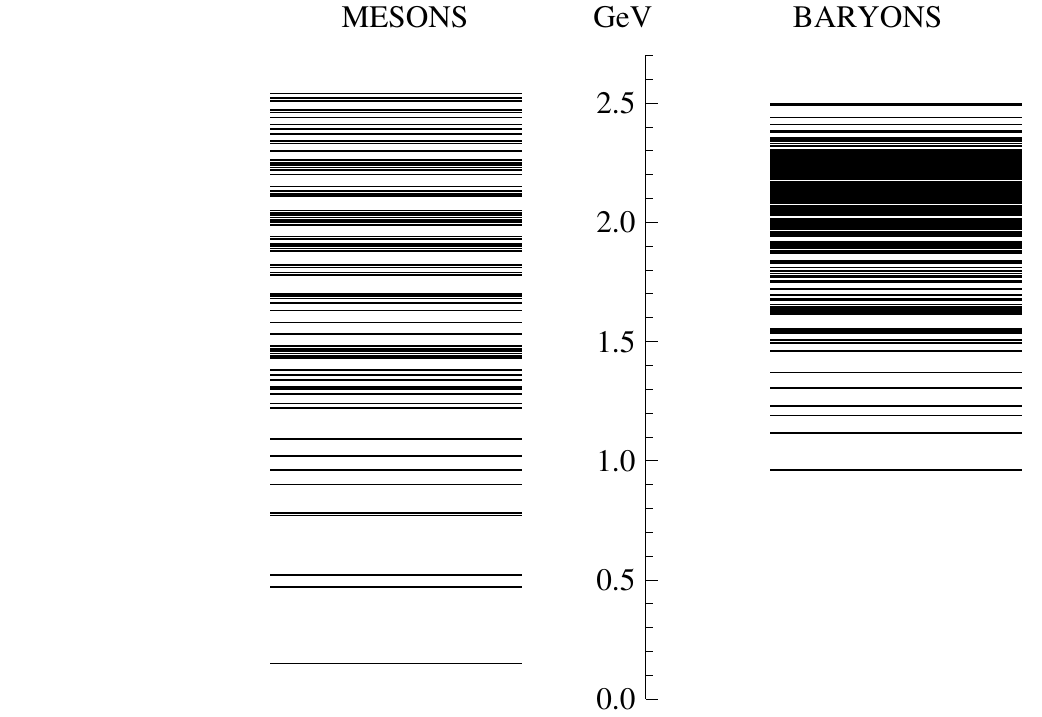,height=5cm,width=6cm,angle=0}
\caption{Left: The full hadron spectrum made of $u,d,s$ quarks from the
  Relativized Quark Model (RQM) of Isgur and collaborators where
  mesons are described as bound $\bar{q} q$ states and baryons as bound
  $qqq$ states~\cite{Godfrey:1985xj,Capstick:1986bm}. Right: Mesonic
  and Baryonic spectrum.}
\label{fig:spectrum-RQM}
\end{figure}

The spectrum of the RQM model of Isgur and
collaborators for $\bar q q$ in the case of mesons and $qqq$ for
baryons~\cite{Godfrey:1985xj,Capstick:1986bm} is
concisely shown in Fig.~\ref{fig:spectrum-RQM}. A detailed comparison
to individual states unveils a rather good description of the data.
Of course, we do not expect this or any quark model Hamiltonian to
describe accurately the individual levels. This, however, poses an
interesting problem on how two different spectra including many
excited states can quantitatively be compared, beyond eyesight and
subjective impression just based on contemplating
Figs.~\ref{fig:spectrum} and \ref{fig:spectrum-RQM}. One way suggested
by Hagedorn in the early 60's is by means of the cumulative number of
states
\begin{eqnarray}
N(M) = \sum_n \theta (M-M_n) \,.
\end{eqnarray}
The question is then to decide to {\it what} extent $N_{\rm QCD}(M)$ or
$N_{\rm PDG}(M)$ coincide.

\begin{figure}
\epsfig{figure=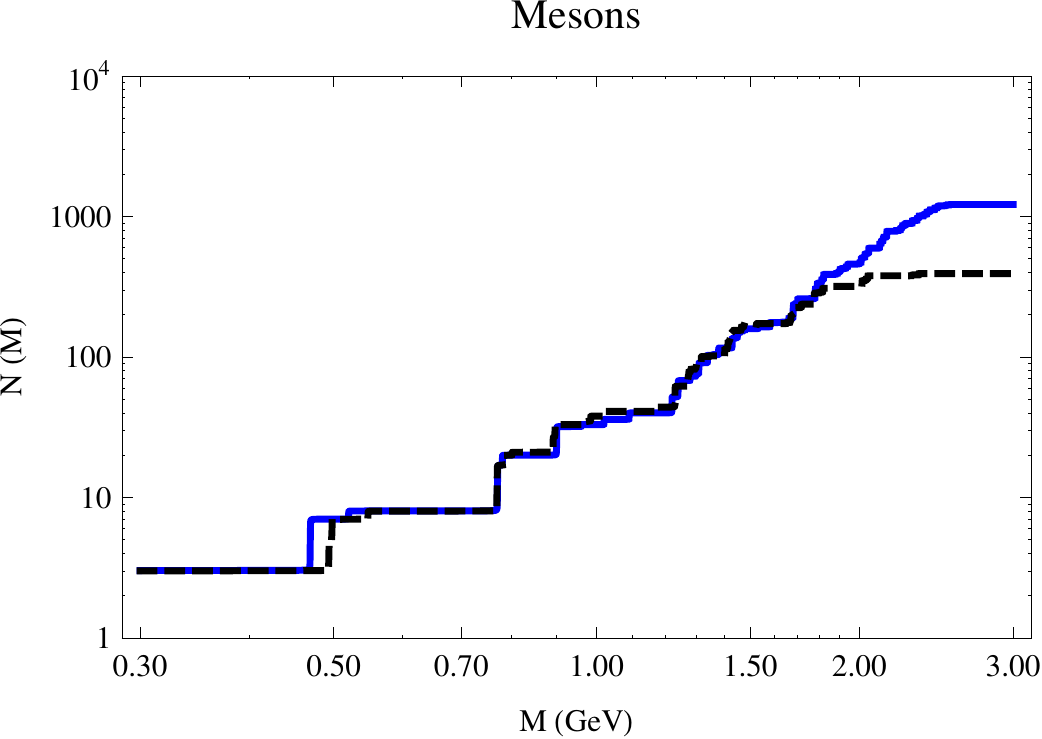,height=7cm,width=6cm,angle=0}
\epsfig{figure=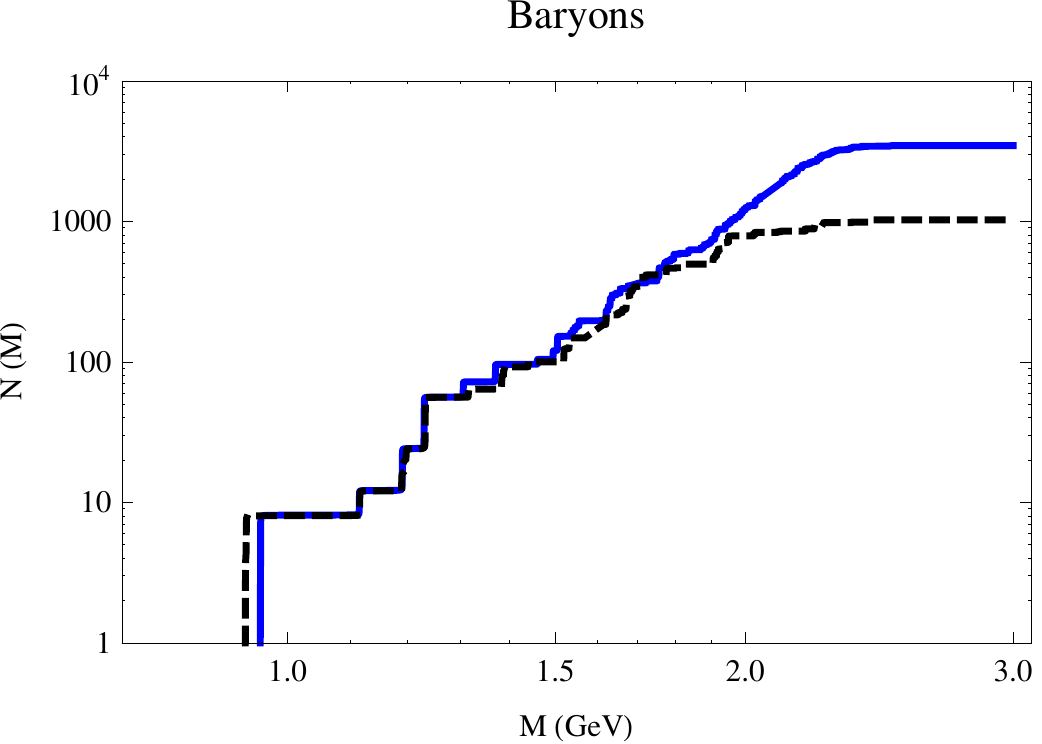,height=7cm,width=6cm,angle=0}
\caption{Cumulative numbers for the PDG (dashed line) and the RQM
  (full line). Left panel: mesonic states (log-log
  scale). Right panel: Baryonic states (log-log scale).}
\label{fig:nucum-meson-baryon}
\end{figure}

For our discussion we will consider a simplified version of that model where
hyperfine splittings due to spin and three-body forces are ignored and the
Hamiltonian for $n$ constituents (we restrict to $q\bar{q}$ and $qqq$ systems)
is taken to be
\begin{eqnarray}
H_n = \sum_{i=1}^n \sqrt{p_i^2+m_i^2} 
- \sum_{i < j} \lambda_i \cdot \lambda_j v(r_{ij}),
\qquad
v(r) = \kappa \, r  - \frac{\alpha_s}{r}
\,.
\label{eq:H-parton}
\end{eqnarray}
We will consider explicitly some cases of interest below, but already
at this level some important observations can be made on the growth of
the cumulative number of states. 
To this end, let us adopt a semiclassical approximation, which should be reliable when the
number of states is large. The number of states in the CM system and at rest,
below a certain mass $M$ takes the form
\begin{eqnarray}
N_n (M) \sim g_n \int \prod_{i=1}^n \frac{d^3 x_i d^3 p_i}{(2\pi)^3}
\delta^{(3)}\left({\textstyle \sum_{i=1}^n \vec x_i}\right) 
\delta^{(3)}\left({\textstyle \sum_{i=1}^n \vec p_i}\right)
 \theta (M-H_n) \,,
\label{eq:ncum-n}
\end{eqnarray}
where $g_n$ takes into account the degeneracy.
For the sake of the argument, let us neglect
the Coulomb term, thus $v(r)=\kappa\,r$, as well as the current quark masses.
In this case, a dimensional argument, $p\to M p$, $r\to Mr/\kappa$, gives  
\begin{eqnarray}
N_n (M) 
\sim \left(\frac{M^2}{\kappa}\right)^{3 n-3}
.
\end{eqnarray}
It is not hard to show that lifting any of the above approximations only
modifies this result by subleading powers of $M$.
Thus, for a finite number of degrees of freedom, the leading contribution to
the cumulative number scales as a power of the mass. The qualitative power
behaviour can be clearly identified as straight lines in the log-log plot of
Fig.~\ref{fig:nucum-meson-baryon} for the cumulative number, where we compare
the resulting cumulative number both for the PDG and the RQM separating the
mesonic and baryonic contributions.

As we can also see, the PDG and RQM spectra look very much alike below $
1.7-1.8 \GeV$. We note, however, that the spectrum for the RQM saturates
sharply at $2.5 \GeV$ which is about the cut-off mass where Isgur, Godfrey and
Capstick stopped to compute states. On the other hand, the PDG states
saturates at lower energy values in a softer fashion.  Note that the RQM looks
like a linear extrapolation (mind the log scale) of the PDG spectrum. The
agreement at lower masses is not highly surprising since RQM parameters were
tuned to reproduce low lying states on the one side {\it and} the listed PDG
states fit into the quark model scheme on the other side. On the other hand,
there are only $\sigma$ and the $u,d,s$ quark masses as fitting parameters, so
we should regard the agreement as a further big success of the RQM from a
global point of view.

\subsection{Bound states vs resonances}

One issue which is problematic from the start is that {\it if} $N(M)$
really counts the number of bound states, then we expect it to be, for
$N_f=2$ flavors, just pions $(\pi^+, \pi^-, \pi^0) $, nucleons
$(p,n)$ and anti-nucleons $(\bar p ,\bar n)$, so $N(\infty)= 3+
2\times2\times2=11$. More generally we have $N (\infty)= (N_f^2-1) +
2\times 2 \times N_f(N_f^2-1)/3$ low lying mesons and baryons, while
in the RQM $N(M)$ diverges as a power.

The cumulative number is by its own stair-case nature a piecewise
discontinuous function, but as we go to higher states the
discontinuities smooth out. This becomes particularly visible in the
RQM in the range $2 \GeV < M < 2.5 \GeV$.

In addition, while in the PDG we have an issue regarding completeness of
states, in the RQM case this is not the case; besides angular and spin
flavor quantum numbers, the radial number can just be checked with
the oscillation theorem, in the $\bar q q$ case, or simply by diagonalizing in
a complete basis of normalizable states; no state will be left out in
the process. In the PDG however, it is unclear if there are missing or
redundant states although the consensus of listing states fitting into
the quark model makes this argument into a circular one.

Of course, the meaning of $M_{n}^{\rm PDG}$ and $M_{n}^{\rm RQM}$ is
different. While in the PDG listing we usually encounter a
Breit-Wigner resonance parameterization characterized by a mass and a
width, in the RQM we have just the mass of a bound $q \bar q $ or
$qqq$ state. We expect that when we couple these bound states obtained
from the RQM to the continuum there will be, besides a width, a mass
shift $M_{n}^{\rm RQM} \to M_{n}^{\rm RQM} + \Delta M_n$, whence the
cumulative number for bound states will generally differ as the one
for Breit-Wigner resonances, but the sign of $\Delta M_n$ is {\it a priori}
unknown. 

\subsection{Hadron sizes}

The solution of the multiparton Hamiltonian Eq.~(\ref{eq:H-parton}) onto
colour singlet states yields the corresponding hadron wave functions. To
estimate the hadron size we can make use of the virial theorem based on
stationarity of eigenstates under unitary coordinate scaling $ \vec r_i \to
\lambda \vec r_i$ around the $\lambda=1$ value, $\Psi (\vec x_1, \dots, \vec
x_n) \to \lambda^{3n/2} \Psi (\lambda \vec x_1, \dots, \lambda \vec x_n)
$. For light quarks we may, for simplicity, take the massless quark limit
$m_q\to 0$. Thus, for the various terms in the Hamiltonian $\langle p_i
\rangle \to \langle p_i \rangle / \lambda $, likewise $\langle 1/r_{ij}
\rangle \to \langle 1/r_{ij} \rangle / \lambda $ and $\langle r_{ij} \rangle
\to \lambda \langle r_{ij} \rangle $ where $\langle r_{ij} \rangle $ is the
mean distance between particles $i$ and $j$.  Due to the virial theorem we
have for a multiparton state
\begin{eqnarray}
M = - 2 \kappa \sum_{i<j} \lambda_i \cdot \lambda_j  \langle r_{ij} \rangle \,.
\end{eqnarray}
In particular, 
\begin{eqnarray}
M_{\bar q q} = 2 \sigma \langle  
r_{\bar q q} \rangle  \, , \qquad  M_{qqq} = \sigma \langle
r_{12}+r_{13}+r_{23} \rangle = 3 \sigma \langle r_{qq} \rangle \,,
\label{eq:mass-size}
\end{eqnarray}
for mesons and baryons respectively. So, in this model the size of a hadron
grows {\it linearly} with the mass per constituent. For instance, from the
relations $m_\rho = 2 \sigma \langle r_{\bar q q} \rangle_\rho $ and $M_N = 3
\sigma \langle r_{q q} \rangle_N $ we have $\langle r_{\bar q q} \rangle_\rho
= 0.42 \fm$ and $\langle r_{q q} \rangle_N = 0.52 \fm$. From $M_N/m_\rho =
3/2$ we would get $ \langle r_{\bar q q} \rangle_\rho = \langle r_{q q}
\rangle_N $ so we can identify the constituent quark mass as $M_q = \sigma
\langle r \rangle $. Note that the above virial relation includes the Coulomb
like contribution $-\alpha_s/r$, since this term scales exactly as the kinetic
piece. In the case of hadrons with one heavy quark, the virial theorem yields
$ M_{\bar q Q} = 2 \sigma \langle r_{qQ} \rangle + m_Q $ and $ M_{q q Q} = 2
\sigma \langle r_{qQ} \rangle + \sigma \langle r_{qq} \rangle + m_Q $.

\subsection{String breaking}

Confinement is often attributed to this ever-linear growing of the energy with
the distance. This is true on the lattice only in the quenched approximation,
where quark-antiquark creation is suppressed. In full QCD however, the string
breaks, a fact that has been observed by lattice calculations
at a distance $r_c=1.25 \fm$  \cite{Bali:2005fu}.  This happens when a light
$\bar q q $ pair is created in between the heavy quark and anti-quark sources
$Q \bar Q$, thus two colour singlet $\bar q Q$ and $\bar Q q$ mesonic
states can be created. On the other hand, charge conjugation implies that the
binding energy of the $\bar q Q$ and the $\bar Q q$ is the same and equals the
residual energy of a heavy-light meson with total mass $M_{\bar q Q}$. Thus
the string breaking distance corresponds to
\begin{eqnarray}
\sigma r_c = 2 \Delta \,, \qquad
\Delta \equiv \Delta_{\bar q Q} =  \Delta_{\bar Q q}
= \lim_{m_Q \to \infty} (M_{\bar q Q}- m_Q) \,.
\end{eqnarray}
Good approximations to these states exist in nature for $q=u,d$ and
$Q=c,b$. From heavy-quark QCD we expect a universal (independent of the heavy
quark spin and flavor) spectrum $\Delta_{h,\alpha}$ of hybrid hadron masses
(heavy-quark mass subtracted). For the lightest, pseudoscalar, hybrid meson,
the following sequence should approach a value of $\Delta$, for increasingly
heavier quarks and using the PDG values in the $\overline{\rm MS}$-scheme,
\begin{eqnarray}
M_K-m_s \equiv \Delta_s &=& 396(24)  \MeV \,,
\nonumber \\
M_D-m_c \equiv \Delta_c &=& 603(81) \MeV \,,
\nonumber \\
M_B-m_b \equiv \Delta_b &=& 1040(130) \MeV \,,
\end{eqnarray}
which gives the estimate for $r_c= 2 \fm$. Another estimate can
be made based on a constituent quark model picture where the total
mass of the quark is $M_q= M_0 + m_q$, with $M_0$ the constituent
quark mass and $m_q$ the current quark mass. Spontaneous breaking of
chiral symmetry implies that in the chiral limit (massless current
quark masses, $m_q \to 0$) the total mass is non vanishing, thus $M_0 \neq
0$.  Actually, for light $u,d$ mesons current masses can be
neglected.  Then the light $\bar q q$ meson has a mass $M_{\bar q q} =
2 M_q$ and the mass of the light $qqq$ baryon is $M_{qqq} = 3 M_q$.
The mass of a heavy-light meson would then be $M_{\bar q Q} = M_q+M_Q
= 2 M_0 + m_q + m_Q $ hence $ \sigma r_c = 4 M_0 + 2 m_q $, which for
$\sqrt{\sigma} = 420-440 \MeV$ and $M_0=300-350 \MeV$ yields
the estimate $r_c=1.2-1.5 \fm$ for the string breaking distance,
a quite reasonable value.

\subsection{Avoided crossings}

\begin{figure}[h]
\epsfig{figure=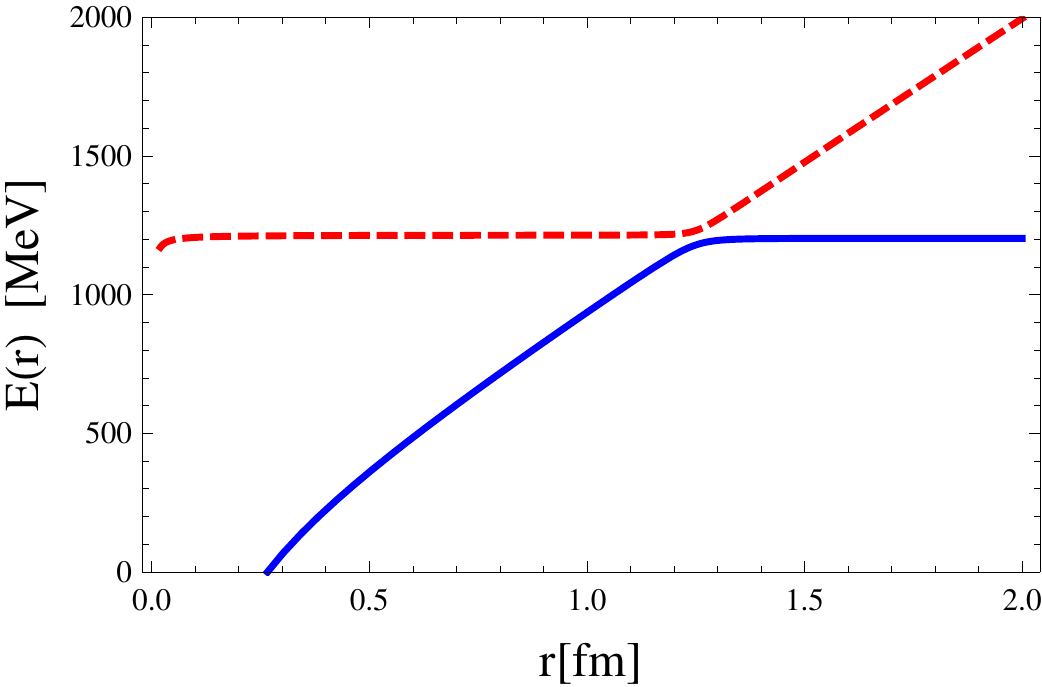,height=5cm,width=6cm,angle=0}
\hskip.5cm
\epsfig{figure=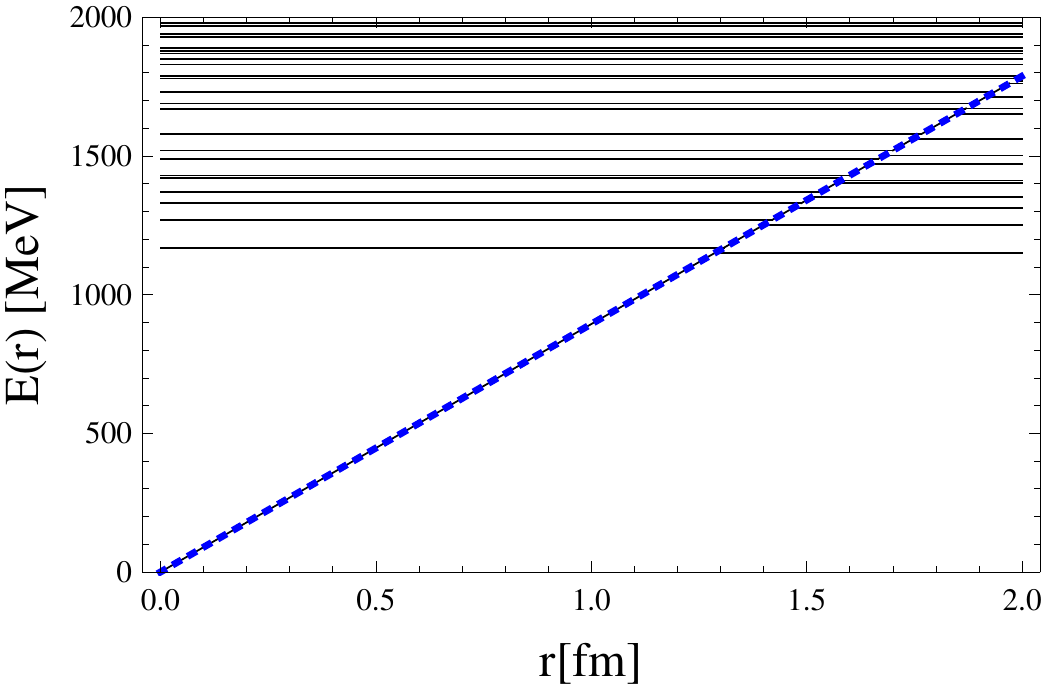,height=5cm,width=6cm,angle=0}
\caption{Avoided crossings structure of adiabatic energy curves (in
  MeV) for heavy $Q \bar Q$ sources as a function of the distance (in
  fm) between the heavy sources. Left panel: Lowest states from
  lattice calculations with a threshold $2 \Delta=1200 {\rm MeV}$
  added~\cite{Bali:2005fu}. Right panel: Complete double heavy-light spectrum
  of the RQM for $c$-hadrons~\cite{Godfrey:1985xj,Capstick:1986bm} and
  included a tiny offset to make clear the structure. We also draw the
  envelope $\sigma r$ with $\sigma= (420 {\rm MeV})^2$.}
\label{fig:Avoided-X}
\end{figure}

The observation of string breaking for a $\bar Q Q$ system requires
taking into account the mesonic $\bar M M$ channels into which the
system may decay after $\bar q q$ pair creation from the vacuum. This
coupled channel dynamics spans the Hilbert space ${\cal H} = {\cal
  H}_{\bar Q Q} + {\cal H}_{\bar Q q \bar q Q} $ and features the
avoided crossing phenomenon familiar from molecular physics in the
Born-Oppenheimer approximation~\cite{landau1965quantum}. For two
channels, say $\bar Q Q$ and $ \bar M M=\bar Q q \bar q Q $, one
can compute the {\it direct} correlators yielding the lowest energies
\begin{eqnarray}
V_{\bar Q Q} (r) = 
\sigma r \, , \qquad V_{\bar Q q \bar q Q} (r) =
\Delta_{\bar q Q}+ \Delta_{\bar Q q} \equiv 2 \Delta \,
. 
\end{eqnarray} 
These two channels are orthogonal for all $r$.  For simplicity we have
disregarded the Coulomb piece $-4 \alpha_s/3 r$ as well as the
residual interaction between the two heavy-light mesons $M=\bar q Q$
and $\bar M=\bar Qq$ which is of van der Waals type and corresponds to
meson exchange. Note that these curves cross when $\sigma r_c = 2
\Delta$. Thus, the spectrum in the Hilbert space with $\bar Q Q$ and
$\bar M M = \bar Q q \bar q Q $ components reads,
\begin{eqnarray}
E_0(r) &=& \sigma r \theta (r_c-r ) + 2 \Delta \theta (r-r_c)  \,,  \\ 
E_0^*(r) &=& 2 \Delta \theta (r_c-r) + \sigma r \theta (r-r_c)  \,,    
\label{eq:string-breaking}
\end{eqnarray}
where now the states $\bar Q Q$ and $\bar Q q \bar q Q $ are piecewise
orthogonal. The point $r=r_c$ corresponding to degenerate states is
singular. A linear combination of $\bar Q Q$ and $\bar Q q \bar q Q $
involves a {\it crossed} correlator between the channels and
representing a variational improvement. The avoided crossing occurs
because the finite energy of the non-diagonal $\bar Q Q \to \bar M M$
interaction lifts the degeneracy, a feature called level repulsion.
The adiabatic potential curves $E_0(r)$ and $E_0^*(r)$ appear as
avoided crossings on the lattice~\cite{Bali:2005fu} with a finite and
small energy repulsion and a narrow transition region of about $0.1
{\rm fm}$, see Fig.~\ref{fig:Avoided-X}, which resemble the simple
shape of Eq.~(\ref{eq:string-breaking}). Therefore, as long as the
size of the system remains small, we may ignore the string breaking
effect. Otherwise, one has to consider a coupled channel dynamics with
$\bar Q Q$ and $\bar Q q \bar q Q$ states.  Excited states potential
curves should follow a similar pattern as
Eq.~(\ref{eq:string-breaking}) but with suitable modifications. Before
mixing one has the crossing among the energy levels up to $\bar q q$
pair creation,
\begin{eqnarray}
V^{(0,0)}_{\bar Q Q} (r) = \sigma r \,, \qquad V^{(n,m)}_{\bar Q q , \bar q Q}(r) = \Delta^{(n)}_{q\bar{Q}}+ \Delta^{(m)}_{\bar{q}Q} \,, 
\label{eq:string-breaking-bis}
\end{eqnarray}
where the double excitation character of the adiabatic potential curve
is displayed explicitly and a universal string tension is assumed. The
crossings must happen at $\sigma r_c^{(n,m)}= \Delta^{(n)}_{q\bar{Q}}+
\Delta^{(m)}_{\bar{q}Q}$. Avoided crossings take place when mixing
among different sectors is allowed yielding energy curves sketched in
Fig.~\ref{fig:Avoided-X} when using the spectrum from the RQM for
$c$-hadrons~\cite{Godfrey:1985xj,Capstick:1986bm}.

\subsection{Limitations in Counting states}

As we have said, when the size of the system is large enough the
string breaks and the assumption of a linear potential becomes
invalid, see Eq.~(\ref{eq:string-breaking}). Because of the linear
dependence of the size with the mass, Eq.~(\ref{eq:mass-size}), this
provides a maximum mass value beyond which the RQM becomes
inapplicable.  A rough estimate for mesons can be made by taking $\max
M_{\bar qq} = 2 \sigma r_c$ which for $r_c=1.25 \fm$ gives $ \max
M_{\bar q q } \sim 2.2 \GeV$. For baryons the situation is more
complex. The value $ \max M_{qqq} = 3 \sigma r_c \sim 3.4 \GeV$ is
actually an upper bound for an equilateral triangular configuration,
however, the string may break more economically when just {\em two}
constituents are sufficiently far apart, $\langle r_{ij} \rangle \sim
r_c$. This corresponds to an elongated isosceles triangle
(quark-diquark configuration) such that $\max M_{qqq} = 2 \sigma r_c +
\sigma \langle r_{12} \rangle \sim 2 \sigma r_c + M_q \sim 2.5
\GeV$. The RQM departs from the PDG at $M_{\bar q q}\sim 2 \GeV$ (see
Fig.~\ref{fig:nucum-meson-baryon}). While this poses the problem on
the validity of the RQM for masses beyond the PDG saturation, it also
suggests that higher mass states break up into weakly bound molecular
systems with a small net contribution to the cumulative
number. Actually, as argued in~\cite{Dashen:1974ns}, counting
hadronic states implicitly averages over some scale, and so states
such as the deuteron generate fluctuations in a smaller
scale.\footnote{The cumulative number in a given channel in the
  continuum with threshold $M_{\rm th}$ is $N(M)= \sum_n \theta(M-M_n)
  + [\delta(M)-\delta(M_{\rm th})]/\pi$ which becomes $N(\infty)=n_B +
  [\delta(\infty)-\delta(M_{\rm th})]/\pi=0$ due to Levinson's
  theorem. In the NN channel where $M_{\rm th}= 2M_N$ the appearance
  of the deuteron changes rapidly at $M= 2 M_N -B_d $ by one unit so
  that $N ( 2M_N-B_d +0^+)- N ( 2M_N-B_d -0^+)=1 $, but when we
  increase the energy this number decreases slowly to zero at about
  pion production threshold $N ( 2M_N+ m_\pi)- N ( 2M_N-B_d -0^+) \sim
  0 $.}

For systems with a heavy quark, we have that for mesons $\Delta_{Q
  \bar q}= M_{Q \bar q}-m_Q = 2 \sigma \langle r_{Q\bar q} \rangle$
and baryons $\Delta_{Q qq}= M_{Q qq }-m_Q = 2 \sigma \langle r_{Q q}
\rangle + \sigma \langle r_{qq} \rangle $ string breaking occurs when
$\Delta_{Q\bar q},\Delta_{Qqq} \sim 2 \sigma r_c$.

\section{Thermodynamics of bound states of quarks}

\subsection{The total cumulative number and equation of state in the confined phase}

The relativized quark model  describes all states as bound
states of $\bar q q$ for mesons and $qqq$ for
baryons~\cite{Godfrey:1985xj,Capstick:1986bm}. The 
total cumulative number is then defined as  
\begin{eqnarray}
N(M)= N_{\bar q q} (M) + N_{qqq} (M) \,. \label{eq:Nrqm}
\end{eqnarray}
This counts the number of bound states below $M$ which is depicted in
Fig.~\ref{fig:cum-all} (note the log-scale) where a clear straight line is
observed. %This suggests a fit of the form
%\begin{eqnarray}
%N(M)= A \, e^{M/T_{\rm RQM}} \,,
%\end{eqnarray}
%for which we get $T_{\rm RQM}=220\MeV$.

%\subsection{The equation of state in the confined phase}

By quark-hadron duality, in the limit of very low temperatures we expect to
have a gas of pions (the lightest hadrons) which due to spontaneous breaking
of chiral symmetry interact weakly at low energies through derivative
couplings. In the chiral limit the pions would become massless resulting in a
small trace anomaly in the temperature regime where heavier hadrons are
suppressed. For a gas of hadrons the pressure reads 
\begin{eqnarray}
P = \sum_n \eta_n \int \frac{d^3 p }{(2\pi)^3} \log \left[ 1+ \eta_n
  e^{-\sqrt{p^2+M_n^2}/T} \right] \,,
\end{eqnarray}
where the sum is over {\it all} hadronic states including spin-isospin
and anti-particle degeneracies. From here, and using the cumulative
number Eq.~(\ref{eq:Nrqm}) obtained with the
RQM~\cite{Godfrey:1985xj,Capstick:1986bm}, it is straightforward to
compute the trace anomaly. The comparison, shown in
Fig.~\ref{fig:trace-PDG}, with the continuum extrapolated results of
the WB~\cite{Borsanyi:2013bia} and HotQCD~\cite{Bazavov:2014pvz}
collaborations is remarkable. As a side remark let us mention that
in QCD $80\%$ of the trace anomaly stems from the gluonic part of the
operator (right-hand of Eq. (\ref{eq:G2}))~\cite{Bazavov:2014pvz}.

\subsection{Polyakov loop correlators}

\begin{figure}
\centering
\epsfig{figure=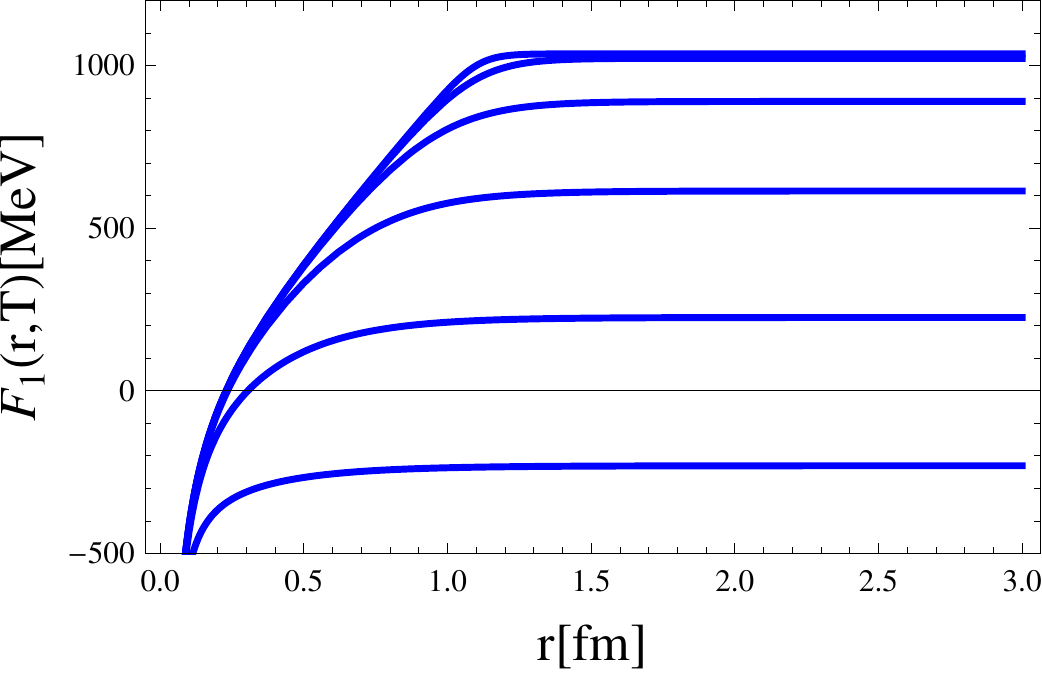,height=5cm,width=6cm,angle=0}
\caption{Free energy $F_1(r,T)$ (in MeV) as a function of the distance
  for a set of increasing temperatures $T = 50, 100, 150, 200, 250,
  300 \, {\rm MeV}$ (from top to bottom), using the RQM in the case of
  c-quarks. We take $\sqrt{\sigma}= 420 {\rm MeV}$.}
\label{fig:F1}
\end{figure}

A straightforward consequence of Eq.~(\ref{eq:string-breaking-bis}) is
that in the confined phase the correlator between Polyakov loops in
the fundamental representation at large distances becomes, according
to Eq.~(\ref{eq:pol-corr-unq}), with $w_0=1$ and $E_0(r)= \sigma r$,
\begin{eqnarray}
e^{-F_1(r,T)/T} &=&\langle {\rm Tr}_F\Omega (\vec r) {\rm Tr}_F \Omega
(0)^\dagger \rangle = \sum_{n,m} e^{- V^{(n,m)}_{\bar Q Q} (r)/T} \nonumber \\ 
&=& e^{-\sigma r/T} + \left(\sum_n e^{- \Delta_n /T} \right)^2  \,,
\label{eq:pol-corr-f}
\end{eqnarray}
where $\Delta_n= \Delta_{\bar{q}Q}^{(n)} = \Delta_{q\bar{Q}}^{(n)}$ by
charge conjugation. Then one has
\begin{equation}
F_1 (r,T) = -T \log\left[ e^{-V_{\bar Q Q}(r)/T}+ e^{-F_1(\infty,T)/T} \right] \,,
\end{equation}
where we have replaced $\sigma r \to V_{\bar Q Q}(r) = \sigma r -
\pi/(12 r)$. The result is depicted in Fig.~\ref{fig:F1} for $T=50,
100, 150, 200, 250, 300 \, {\rm MeV}$ using the RQM in the case of
c-quarks for $F_1(\infty,T)$. We see that at small temperatures there
is little change in qualitative agreement with lattice
calculations~\cite{Kaczmarek:2005ui}.

Corrections to Eq.~(\ref{eq:pol-corr-f}) are expected, as it is based
on a sharp string breaking transition and the fact that we truncated
the spectrum to one light $\bar q q$ pair creation.  The avoided
crossing structure shown in Fig.~\ref{fig:Avoided-X} is modified by
the finite string breaking transition region, which on the lattice and
for the ground state was found to be about $\Delta r=0.1 {\rm fm}$.
Moreover, for $\Delta_{\bar q Q}= \sigma r_c$ we expect the $\bar q Q
$ system to break up into $\bar q Q$ and $\bar q q$.

\subsection{The Polyakov loop}

\begin{figure}
\epsfig{figure= 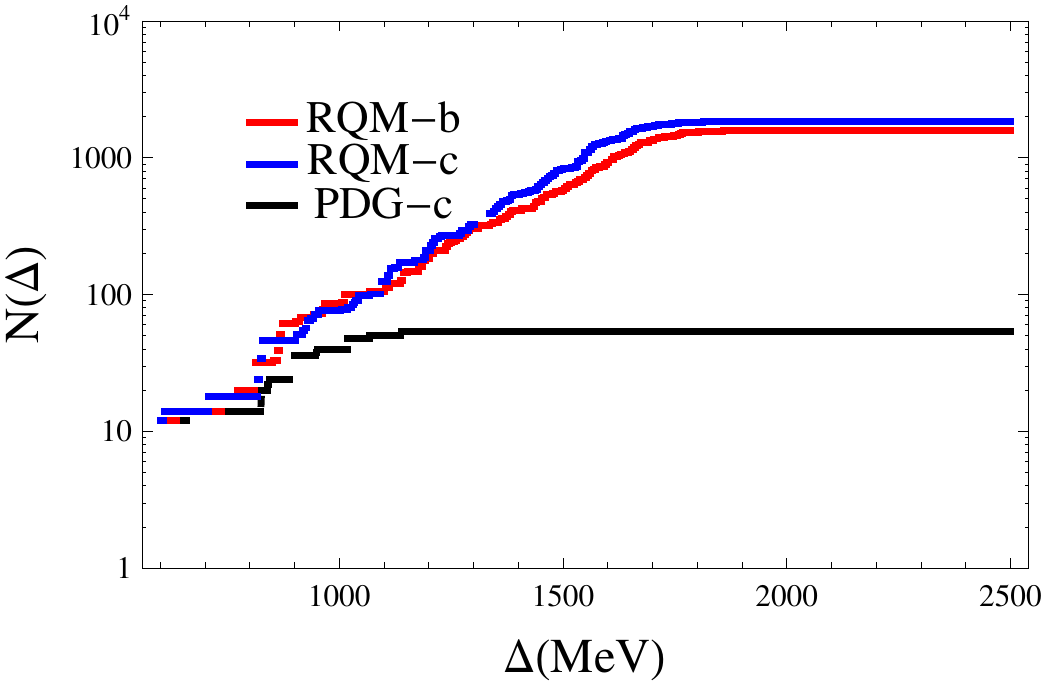,height=5cm,width=6cm,angle=0}
\epsfig{figure=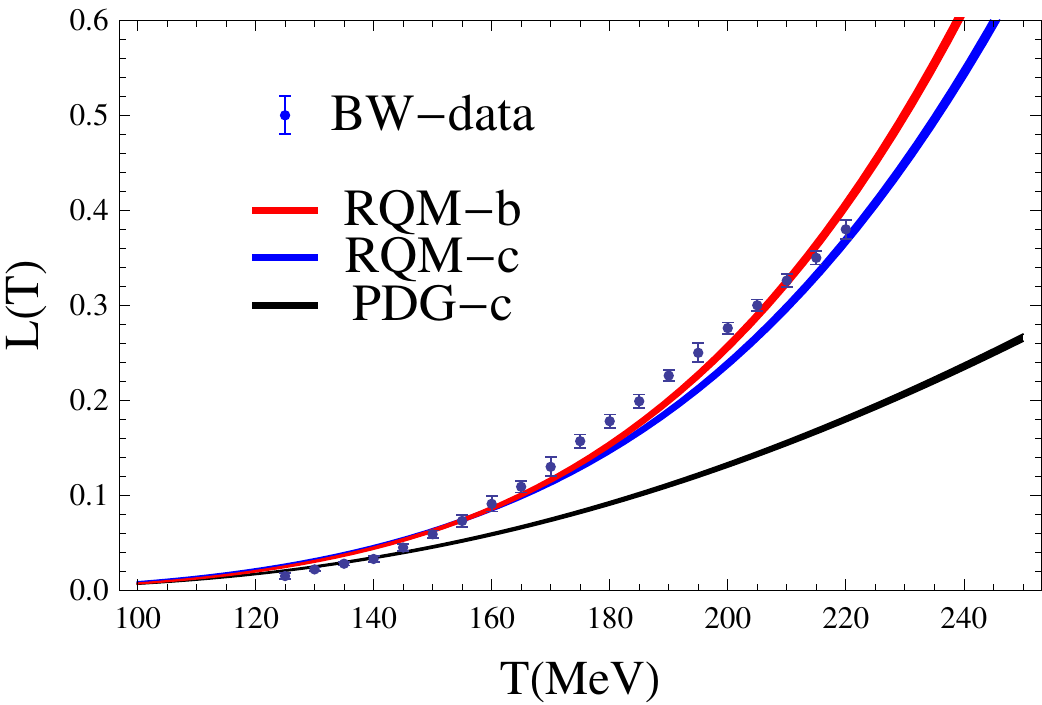,height=5cm,width=6cm,angle=0}
\caption{Left: Cumulative number $N(\Delta)$ as a function of the
  $c$-quark and $b$-quark mass subtracted hadron mass $\Delta= M-m_Q$
  (in $\MeV$) with $u$, $d$ and $s$ quarks, computed in the
  RQM~\cite{Godfrey:1985xj,Capstick:1986bm} and from the
  PDG~\cite{Nakamura:2010zzi}. Right: Polyakov loop as a
  function of temperature (in MeV). Lattice data
  from~\cite{Bazavov:2011nk} for the HISQ/tree action
  and~\cite{Borsanyi:2010bp} for the continuum extrapolated stout
  result. We compare lowest-lying charmed hadrons from
  PDG~\cite{Nakamura:2010zzi}, the RQM spectrum with one $b$ or one
  $c$-quark~\cite{Megias:2012kb}. }
\label{fig:heavy}
\end{figure}

We can likewise consider the spectrum of a system with one heavy quark
such as $c,b,t$ quarks. An equivalent cumulative number can also be
defined with similar features. Because of the heavy mass it is more
convenient to subtract the heavy quark mass, $m_Q$, from the hadron
mass, $M= \Delta + m_Q$. Using again the RQM for hadrons with one
heavy quark, $\bar q Q$ for mesons and $Qqq$ for
baryons~\cite{Godfrey:1985xj,Capstick:1986bm}, the total cumulative
number is defined as
\begin{eqnarray}
N(\Delta)= N_{\bar q Q} (\Delta) + N_{Qqq} (\Delta) \,.
\end{eqnarray}
The result is depicted in Fig.~\ref{fig:heavy}, where similar patterns
as the light quark systems are encountered, namely power behaviour for
large $\Delta$ for individual $N_{\bar q Q} (\Delta) $ and $ N_{Qqq}
(\Delta)$ contributions, and Hagedorn type spectrum for the combined
result, with $T_H \sim 210 {\rm MeV}$~\cite{Arriola:2013jxa}.

The Polyakov loop is obtained by computing the corresponding partition
function of $N_{\bar q Q} (\Delta) $ and $ N_{Qqq} (\Delta)$ with the
pertinent statistics since from Eq.~(\ref{eq:pol-corr-loop}) and
Eq.~(\ref{eq:pol-corr-f}) we get
\begin{equation}
L_T = \sum_n e^{-\Delta_n/T} \,.
\end{equation}
The result can be seen in Fig.~\ref{fig:heavy} and compared to lattice
data from~\cite{Bazavov:2011nk} for the HISQ/tree action
and~\cite{Borsanyi:2010bp} for the continuum extrapolated stout
result.\footnote{We are cavalier on the renormalization issues and
  multiplicative ambiguities in $L_T$. Further details are discussed in
  Ref.~\cite{Megias:2012kb}.}

\section{Limitations of the Hadron Resonance gas model}

\subsection{Hagedorn spectrum}

As we have mentioned, the HRG has been successfully applied in many
situations below the phase transition, such as the EoS, and quark
number susceptibilities, but little has been achieved with regard to
understanding the emergence of hadronization in the low temperature
regime. So, what is the complete hadron spectrum? Of course, we
expect QCD to give the answer to this question, but this requires a
knowledge of all multiparticle states of stable particles, and most of
them are in the continuum. The PDG tries to answer this question by
filling in the expected quark model states $\bar q q$ for mesons and
$qqq$ for baryons. While states falling outside this category are
listed as further states, whether or not the list contains redundant
states is difficult to say.

\begin{figure}[h]
\begin{center}
\epsfig{figure=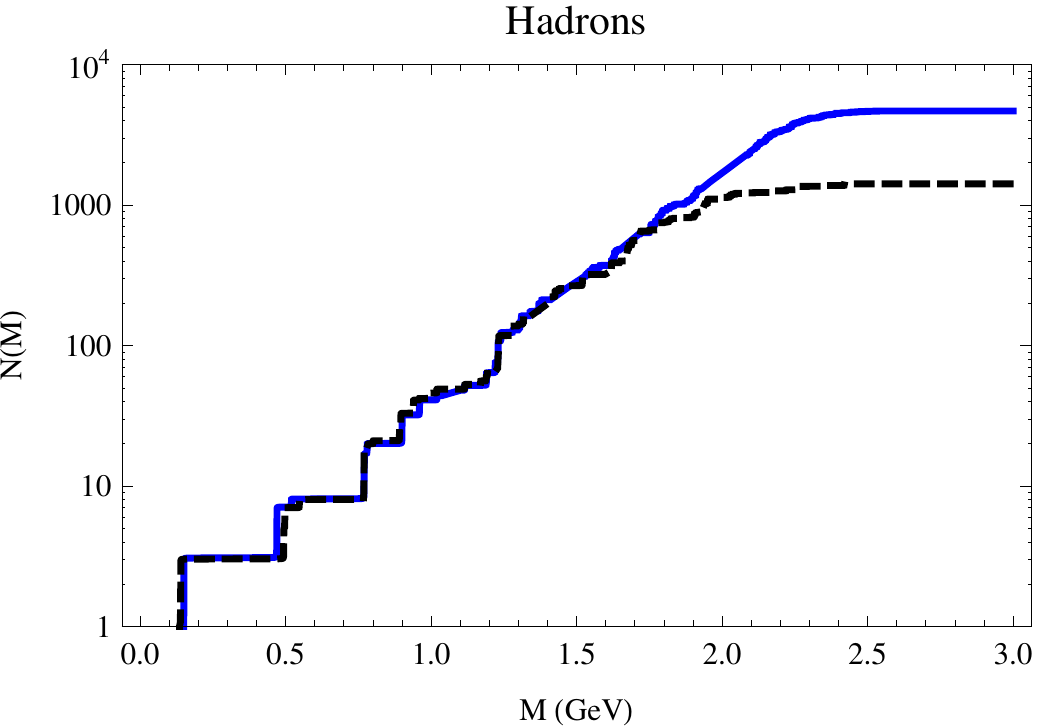,height=5cm,width=6cm,angle=0}
\end{center}
\caption{Cumulative number for the PDG~\cite{Nakamura:2010zzi} (dashed) and the RQM~\cite{Godfrey:1985xj,Capstick:1986bm} (full).}
\label{fig:cum-all}
\end{figure}

In Fig.~\ref{fig:cum-all} we show the total cumulative number for the
PDG~\cite{Nakamura:2010zzi} compared to the
RQM~\cite{Godfrey:1985xj,Capstick:1986bm}. As can be seen there is an
exponential growth for $M \le 2 \GeV$ for the PDG
states~\cite{Nakamura:2010zzi} and $M \le 2.5 \GeV$ for the RQM
states~\cite{Godfrey:1985xj,Capstick:1986bm}.  This remarkable feature
of the cumulative number, first noted by
Hagedorn~\cite{Hagedorn:1965st}, has fascinated theoreticians for
decades. The exponential growth is of the form $ N(M) \sim e^{M/T_H}$
where $T_H$ is the so-called Hagedorn temperature. This pattern was
predicted with very few states and an ever increasing exponential
spectrum was anticipated as new states entered the hadronic list.
Different upgrading analyses have confirmed this exponential
growth~\cite{Broniowski:2000bj,Broniowski:2004yh} and even two
different Hagedorn temperatures have been reported (see however the
discussion around Fig.~\ref{fig:nucum-meson-baryon}). The onset of the
Hagedorn spectrum has been questioned more
recently~\cite{Cohen:2011cr}.  The growth happens at about $M \sim 1.5
\GeV$ and continues with the {\it same} slope until $M \sim 2.1
\GeV$. The pattern is even more clear in the RQM, where the trend
stretches up to $M \sim 2.5 \GeV$. The consequence of a truly Hagedorn
temperature is that the partition function, and hence all
thermodynamics quantities develop a pole at the Hagedorn temperature,
\begin{eqnarray}
\DeltaA_{\rm HRG} \to \frac{A}{T-T_{H}} \,.
\end{eqnarray}
This form is not observed on the lattice although good fits 
in the range $ 50  \MeV < T < 180  \MeV$ for the trace anomaly yield a value of
about $T_H \sim 220 \MeV$. 

One important consequence of such a favorable comparison between PDG
and RQM is that it may provide a clue to the onset of the Hagedorn
spectrum. We can estimate the asymptotic behaviour for both $\bar q q
$ and $qqq$ spectra in the RQM as the model Hamiltonian is known. As
we have shown we expect a power behaviour for both $N_{\bar q q} (M) $
and $N_{q q q} (M) $, a fact confirmed in
Fig.~\ref{fig:nucum-meson-baryon} where the separated contributions in
a log-log scale exhibit this power like behaviour. The question is on how is it possible that summing two
polynomials do we end up with an exponential?

Using the MIT Bag model~\cite{Johnson:1975zp} this question was
answered by Kapusta ~\cite{Kapusta:1981ay} since then the cumulative
number can be computed as $\sum_n N_n(M)/n! g^n $ from
Eq.~(\ref{eq:ncum-n}) by using a constant Bag volume and evaluating the
phase-space integral for free particles. We have checked this
explicitly by summing the Bag modes~\cite{Megias:2013xaa}.

\subsection{Finite width corrections}

Of course, the discrete summation involved in the cumulative number
definition requires considering all masses of states to be interpreted
as bound states. In reality most of the states listed in the PDG are
resonances, i.e. unstable particles which have instead a mass spectrum
characterized by a distribution $\rho(\mu)$ which is peaked at
$\mu=M_n$ with a certain width $\Gamma_n$. 

This idea has been implemented by the half-width rule, where resonance
masses are regarded as random variables with an uncertainty of half
the width~\cite{Arriola:2011en}. This interpretation has been
fruitful in describing the Regge spectrum of
mesons~\cite{Masjuan:2012gc,Masjuan:2013xta}, hadronic form
factors~\cite{Masjuan:2012sk} as well as providing an error estimate for
the cumulative number itself~\cite{Arriola:2012vk}. This also provides
a quantitative way of defining a figure of merit for the quark model
taking the width as a genuine uncertainty.

In a quantum mechanical picture, where the resonance decay can be
viewed as a tunneling process, the mass shift of a unstable state is
negative as the infinite barrier becomes finite making the energy
shift negative. As a consequence, the cumulative number would increase
for a fixed mass, $N_{\rm resonance} (M) > N_{\rm bound} (M)$. This
effect goes in the opposite direction of making $N_{\rm RQM}(M)$ and
$N_{\rm PDG}(M)$ to agree in the upper part of the spectrum, and that
the outnumbering of $\bar q q $ and $qqq$ states in the RQM would be a
genuine one.

\subsection{Excluded volume  condition}

The reason for the failure of the HRG in describing the trace anomaly at $T
\sim 170 \MeV$ may be sought by questioning any of the assumptions
involved. One of them, the finite size of hadrons may be tested by computing
the excluded volume.

In an ideal world where the hadrons live forever they still have a finite
size.  This is the case for instance in the large $N_c$ limit, where
one has $\Gamma/M = {\cal O}(N_c^{-1})$ but their size is $ r = {\cal
  O} (N_c^0)$. Due to the finite hadronic size, when hadrons overlap
their underlying composite nature becomes relevant, particularly
regarding the Pauli principle as applied to the constituent quarks. 

The finite volume corrections have often been addressed along the lines of
statistical mechanics for real gasses where hadron volumes are usually assumed
to be similar (see e.g. \cite{Begun:2012rf}). Taking into account how these
corrections originate in the quantum virial expansion as repulsive
contributions through negative phase shifts it is unclear what are the actual
values one should take for the volume without actually carrying the phase
shift analysis.  In a bound states picture, which is the spirit of the HRG
since finite width effects are disregarded, the size of the bound state is
expected to increase with the mass of the state.

One simple estimate of the size of the hadron can be made by using the
MIT bag model~\cite{Johnson:1975zp} where the volume $V_i$ of the
hadron is a natural concept since in that model hadrons with a mass
$M_i$ have a sharp edge, $r_i$, where
\begin{eqnarray}
V_i = \frac{4\pi}{3} r_i^3 = M_i/(4B)  \,, \qquad B=(0.166 \GeV)^4 \,, 
\label{eq:vol-BAG}
\end{eqnarray}
and thus the volume grows with the mass. 

In the case of the RQM where particles are interacting through a
confining $\sigma r_{ij}$ potential, and hadrons are localized but
have a diffuse edge, the volume is not a well defined concept. 
To estimate the value of the volume occupied by the hadron in the RQM,
we take as the meson radius $\langle r_{\bar{q}q} \rangle/2$ and for
the baryon the radius of the equilateral triangle $\langle r_{qq}
\rangle /\sqrt{3}$.  This yields the volume estimate
\begin{eqnarray}
V_i = \frac{4\pi}{3}\frac{M_i^3 c_i^3}{\sigma^3}  \, ,  \qquad 
\sqrt{\sigma}= 0.42 \GeV ,  
\label{eq:vol-RQM}
\end{eqnarray}
where $c_i = 1/4$ for mesons and $c_i = 1/(3\sqrt{3})$ for baryons. With this
prescription, baryons occupy a smaller volume for the same mass. These are
crude estimates, which assume a sharp edge of the hadron, but they show that
the volume does depend on the mass.

Rather than trying to model finite volume corrections, we will analyze
the quite natural condition that the excluded volume cannot be
negative, or equivalently that the occupied volume is smaller than the
total volume. This means
\begin{eqnarray}
\sum_i V_i N_i \le V \,, \qquad \sum_i  V_i \int \frac{d^3 p}{(2\pi)^3} \frac{g_i}{e^{E_i(p)/T} \pm 1} \le 1 \,. 
\end{eqnarray}
\begin{figure}
\begin{center}
\epsfig{figure=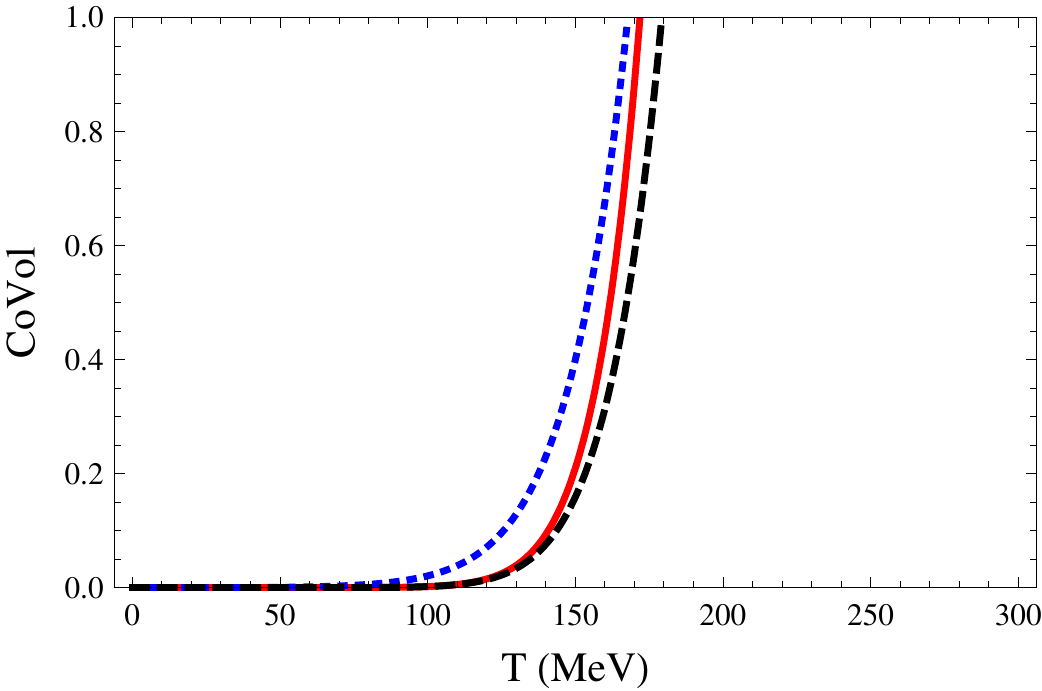,height=5cm,width=6cm,angle=0}
\epsfig{figure=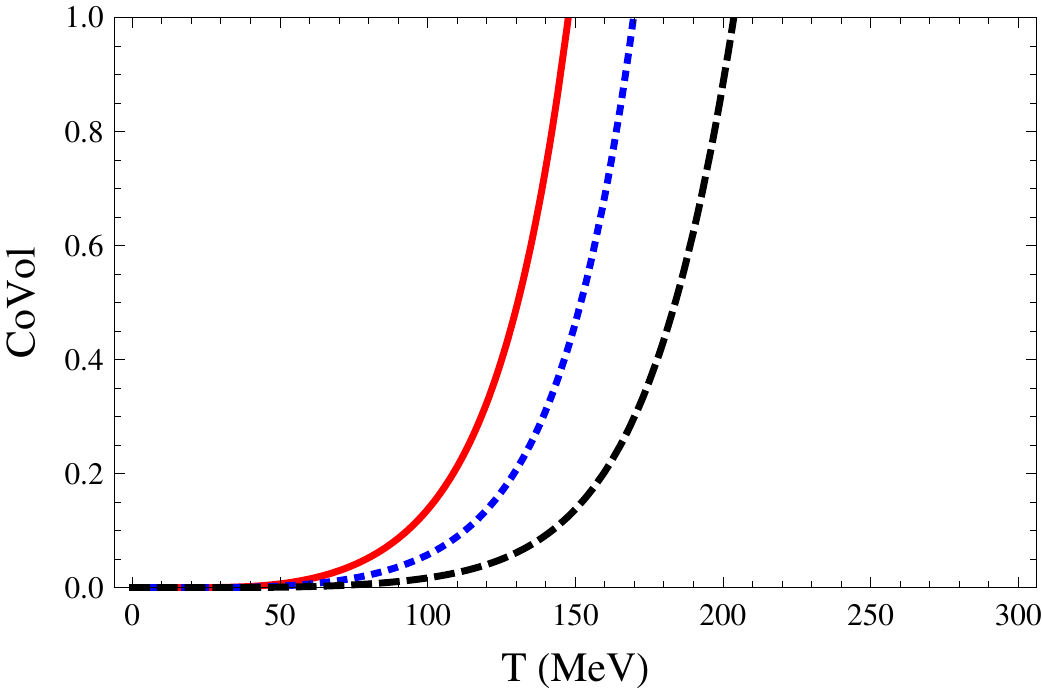,height=5cm,width=6cm,angle=0}
\end{center}
\caption{The occupied volume condition using three different models
  for the relation between the volume and the mass for a hadron. The
  upper edge of the figure expresses the maximum limit beyond which
  the conditions is violated. We show: (Left panel) RQM spectrum with
  $V_i \sim M_i^3$ (Dotted), PDG spectrum with $V_i = M_i/(4B)$ (Full),
  PDG spectrum with $V_i = M_i^3$ (Dashed). (Right panel) PDG
  spectrum with constant volume using $r=1 \fm$ (full), $r=0.75
  \fm$ (dotted) and $r=0.5 \fm$ (dashed).}
\label{fig:covol}
\end{figure}
In Fig.~\ref{fig:covol} we illustrate the situation by depicting the
occupied volume condition using the MIT Bag model volume,
Eq.~(\ref{eq:vol-BAG}), and our estimate for the RQM hadronic volume,
Eq.~(\ref{eq:vol-RQM}), using both the RQM spectrum and the PDG
spectrum. As one can see the condition is violated for temperatures $T
\sim 150-180 \MeV$ where the HRG departs from the lattice data. This
is an interesting fact, since it relates the crossover to a volume
effect, which deserves further investigation.  For comparison we also
show in the right panel the results for a constant hadronic volume
taking the radius values $r=0.5,0.75,1 \fm$. The choice $r \sim 0.75
\fm$ resembles the RQM and bag excluded volume condition.

\section{Insights from gluodynamics}

Matters get simpler in the case of pure gauge theories where gluons
are the only degrees of freedom. In QCD this corresponds to taking the
limit $m_q \to \infty$ for {\it all} quark species.  As mentioned, the
Polyakov loop in the {\it fundamental} representation becomes a true
order parameter of the deconfinement phase transition since at low
temperatures it vanishes as $L_T= {\cal O} (e^{-m_q/T})$ in this
limit. The Polyakov loop in the {\it adjoint} representation does not
vanish below $T_c$. Gluon-Glueball duality means in this case that
{\it any} observable defined as an expectation value of a gauge
invariant and hence colour singlet operator can be determined in terms
of purely colour singlet states in the confined phase. As we will
illustrate, this means in practice that the EoS can be determined in
terms of glueballs (bound states of gluons) whereas the Polyakov loop
in the adjoint representation can be be determined from gluelumps
which are bound states of gluons in the presence of a colour octet
source~\cite{Karl:1999wq,Simonov:2000ky,Guo:2007sm,Marsh:2013xsa}.

\subsection{The glueball gas model}

In pure gauge theories the glueball spectrum has been determined in
the lattice~\cite{Meyer:2004gx}. (For full unquenched QCD glueball
studies see e.g.~\cite{Gregory:2012hu}. A general perspective
including phenomenology has been reviewed in
Ref.~\cite{Mathieu:2008me}.)

Gluons are massless particles for which the helicity
formalism~\cite{Jacob:1959at} is the proper one.  For spin 1 particles
this means that just the $\pm 1$ projections are possible. This issue
was first discussed by Barnes~\cite{Barnes:1981ac} within the context
of gluonium (see \cite{Mathieu:2008bf,Buisseret:2009ea,Mathieu:2009cc}
for an upgraded discussion).  In the helicity formalism two-gluon
states $J^{PC}=1^{-+}, 1^{++}, 3^{-+},5^{-+}$ are forbidden in
accordance to lattice results~\cite{Morningstar:1999rf,Meyer:2004gx}.
The glueball spectrum of two gluons might be obtained from a full
fledged solution of the Bethe-Salpeter equation for two spin-1
particles~\cite{Szczepaniak:1995cw,Szczepaniak:2003mr,Meyers:2012ka,Swanson:2013hta}. In
practice, solving these equations requires an ansatz for the kernel
which is usually obtained by making reasonable approximations.  For
our discussion we will make some drastic approximations which actually
illustrate one important point regarding gluon-glueball duality at
finite temperature. We will assume the following rotational invariant
and spin independent Hamiltonian in the CM system
\begin{eqnarray}
H \psi_n = \left(2 p + \sigma_A r -\frac{3\alpha_s}{r}\right) \psi_n = M_n \psi_n
\,.
\label{eq:H-2g}
\end{eqnarray}
In this scheme the total spin of the glueball $J$ is obtained by composing the
gluons spin and the relative orbital angular momentum.  Due to Bose statistics
the total glueball wave function must be, besides a colour singlet state,
fully symmetric. Thus, the spin and orbital part must be both either symmetric
or antisymmetric.  This gives a total degeneracy of $ (2l+1) g(g\pm1)/2$ for
even/odd $l$, where $g$ is the number of spin states of the gluon.  The lowest
one is the $0^{++}$ glueball~\cite{West:1995ym}. Then the cumulative number becomes 
\begin{eqnarray}
N(M) = \sum_{n,l} \nu_l (2l+1) \theta (M-M_{nl}) \,,
\end{eqnarray}
where $\nu_l= g(g\pm1)/2$ for even/odd $l$.

A simple estimate using the uncertainty principle for the ground state is made
by taking $p r \sim 1$, thus
\[
M_0 = \min \left[ \frac2{r} + \sigma_A r  -\frac{3\alpha_s}{r} \right] = 2
\sqrt{(2-3\alpha_s) \sigma_A} \approx 4 \sqrt{\sigma} 
\,.
\]
For excited states, an estimate can be obtained by using the Bohr-Sommerfeld
quantization condition (WKB spectrum). For $s$-wave, this gives
\[
2 \int_0^a dr \, p(r) = 2 \pi (n+\alpha) 
,
\]
where $p(r)= (M-\sigma_A r + 3\alpha_s/r)/2$, $a$ is the classical
turning point, $p(a)=0$, and $\alpha$ depends on the boundary
conditions. This produces (neglecting the Coulomb term)
$$
M_n^2 = 4 \pi \sigma_A (n+\alpha)  \qquad \text{(WKB, $s$-wave)}
.
$$

More systematically, the eigenvalues of the CM Hamiltonian,
Eq.~(\ref{eq:H-2g}), can be computed by diagonalizing it in the harmonic
oscillator wave functions basis
\begin{equation}
R_{nl}(r)= \frac{u_{nl}(r)}r = 
\frac{e^{-\frac{r^2}{2 b^2}}}{\sqrt[4]{\pi}} \left(\frac{r}{b}\right)^{l} \sqrt{\frac{(n-1)! 2^{l+n+1}}{b^3 (2 l+2 (n-1)+1)\text{!!}}}
   L_{n-1}^{l+\frac{1}{2}}\left(\frac{r^2}{b^2}\right)  \,,
\end{equation}
where $L_{n-1}^{l+\frac{1}{2}}(x)$ are associated Laguerre
polynomials.  The advantage of the harmonic oscillator basis is that
the matrix elements of the pseudodifferential operator $|\vec{p}|$ are
simply related to those of $r$.~\footnote{Beware that the Fourier-Bessel
  transform introduces an additional phase $(-1)^n$ to the momentum
  space wave function $P_{nl}(p) \equiv \int_0^\infty j_{l} (pr) R_{nl}(r) r^2 dr = (-1)^n R_{nl} (r=p b^2)$.}

The reduced wave functions $u_{nl}(r)$ are normalized to unity
\begin{eqnarray}
\int_0^\infty dr r^2 R_{nl}(r)^2 =  \int_0^\infty dr u_{nl}(r)^2 =  1 \,, \nonumber 
\end{eqnarray}
and satisfy the equation   
\begin{eqnarray}
-u_{nl}''(r) + \left[ \frac{r^2}{b^4} + \frac{l(l+1)}{r^2} \right] u_{nl}(r)=
 \frac{1}{b^2}( 2 l + 4 n -1 ) u_{nl}(r) \,.
\end{eqnarray}
Here $b$ has dimensions of length. The single-particle energies are 
\begin{eqnarray}
\epsilon_{nl} = \frac{1}{2 M b^2}\left( 4 n + 2 l -1 \right)
= \omega \left( 2 n + l -1/2 \right) \,, \nonumber 
\end{eqnarray}
where the oscillator frequency is $\omega=1/(M b^2)$. Thus, we can compute the
matrix elements $H_{nl,n'l}$ up to some maximum values of $n$ and $l$ and
diagonalize the truncated Hamiltonian. A list of eigenvalues (for a linear
potential) for $n \le8$ and $l\le 13$ can be looked up in
Ref.~\cite{Bicudo:2007wt}. The quoted accuracy is right provided we take the
dimension as large as $N=100$.

\begin{center}
\begin{figure}
\epsfig{figure=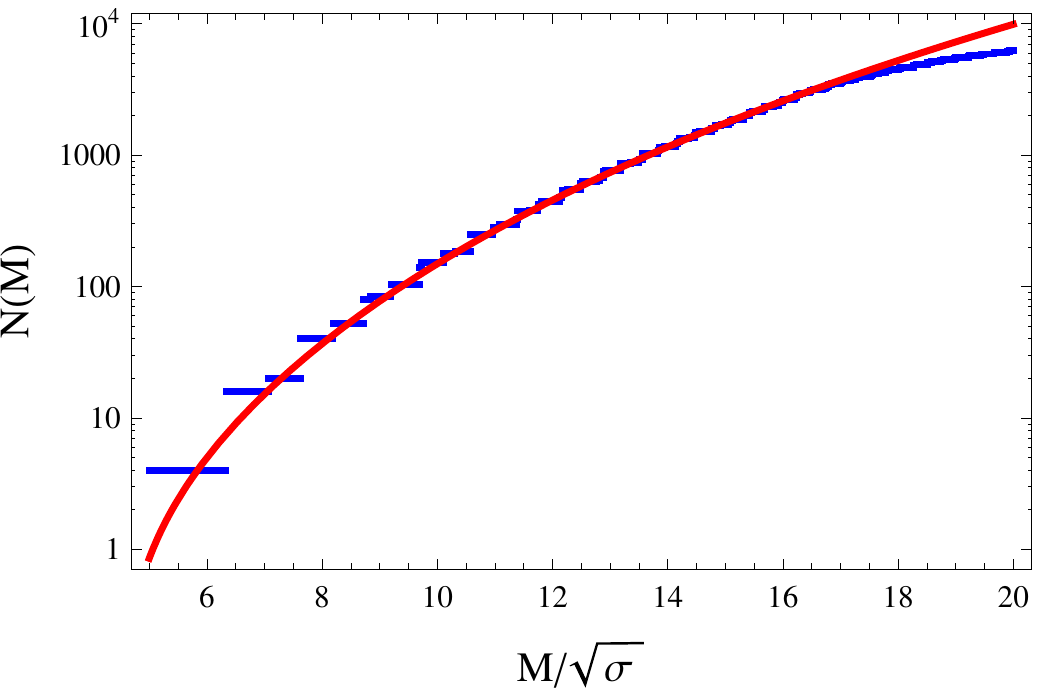,height=5cm,width=6cm,angle=0}
\epsfig{figure=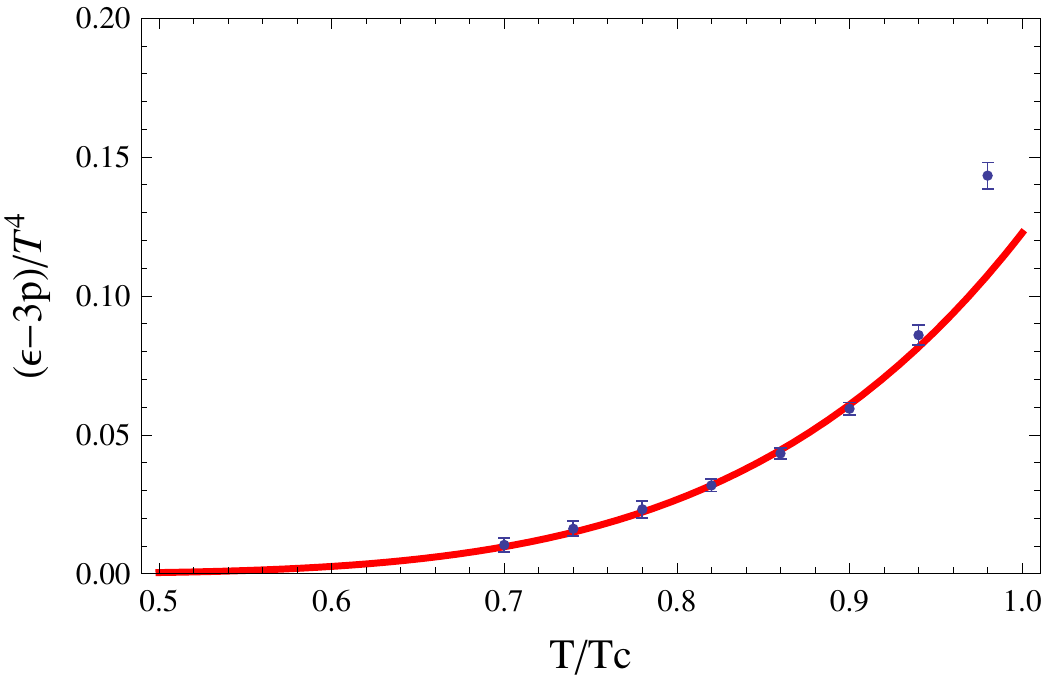,height=5cm,width=6cm,angle=0}
\caption{Left panel: Cumulative number of states for two-gluon
  glueballs as a function of the mass in units of the fundamental
  string tension $\sqrt{\sigma}$, compared to the semiclassical
  approximation. Right panel: Two-gluon states contribution to the
  trace anomaly compared to the lattice data from the WB
  collaboration~\cite{Borsanyi:2012ve}.}
\label{fig:ncum-glue-tran}
\end{figure}
\end{center}

A derivative expansion \cite{Caro:1994ht} can be used to evaluate the
cumulative number of the Hamiltonian in Eq.~(\ref{eq:H-2g}). In the
present context this is closely related to a semiclassical expansion and a
large mass expansion. A direct application of the results in \cite{Caro:1994ht}
gives\footnote{For simplicity here we disregard the effect of the different
  degeneracies in even and odd partial waves. It has no effect at leading
  order in $\hbar$.}
\begin{eqnarray}
N_{2g}(M) &=& \frac{g^2}{2} \int \frac{d^3 x d^3 p}{(2\pi)^3} \theta(M- H(p,r)) + \dots 
\nonumber \\ 
&=& 
\frac{g^2}{2} \left(
\frac{M^6}{720 \pi\sigma_A^3} 
+\frac{\alpha_s M^4}{16 \pi  \sigma_A^2} + \frac{9 \alpha_s^2 M^2}{8 \pi
  \sigma_A }-\frac{M^2}{9 \pi  \sigma_A} +\cdots \right)
.
\end{eqnarray}
By restoring the $\hbar$'s in the Hamiltonian, $\sigma_A \to
\sigma_A/\hbar$ and $\alpha_s\to \hbar \alpha_s$, we can see that the
first quantum corrections enter in the last term quoted. As
advertised, the derivative expansion corresponds to a power expansion
in $M$, with a general form
\begin{eqnarray}
N(M) = \sum_n a_n M^n 
\,.
\label{eq:largeM}
\end{eqnarray}
We show in Fig.~\ref{fig:ncum-glue-tran} the comparison of the full
diagonalization in the Harmonic Oscillator basis to the semiclassical
large mass expansion. As we see, the agreement is rather good for a
relatively low value of $N(M)$.

It is now straightforward to compute the corresponding contribution to the
trace anomaly from the expression\footnote{This expression follows from
the  expansion of Eq.~(\ref{eq:tran-HRG}) in powers of $e^{-E_n(p)/T}$ prior to
  integrate over the momentum.}
\begin{eqnarray}
\DeltaA_{\rm glueball}^{2g}(T)= \sum_{n} \sum_{k=1}^\infty
\frac1{2\pi^2 k} \left(\frac{M_{n}}{T}\right)^3  K_1 \left( \frac{k M_{n}}{T} \right) \,,
\end{eqnarray}
where $M_n$ are eigenvalues of $H$ with the corresponding
spin-orbit-colour degeneracies compatible with the colour singlet and
Bose character of the two-gluon states. The most recent lattice data
from the WB collaboration~\cite{Borsanyi:2012ve} include 7 points for
$\DeltaA (T)$ below $T_c$ which being a dimensionless quantity depends
just on the dimensionless ratio $T/T_c$. In our case we only have the
dimension $\sqrt{\sigma}$ and thus the dimensionless $\DeltaA_{2g} (T)
$ depends on $T/\sqrt{\sigma}$. Thus, we can determine $T_c/
\sqrt{\sigma} $ from a fit of $\DeltaA_{2g}(T)$ to the lattice
data~\cite{Borsanyi:2012ve}.

For $g=3$ we get $\chi^2=3.2$ for the first 6 lattice data but
$\chi^2/\nu=56/(7-1)$ when the last point is added. Actually, the
value of $\sigma $ is already determined by the first point, although
its error decreases as new points are included in the fit. This
provides some robustness to the analysis, and indicates that for $T
\le 0.9 T_c$ the trace anomaly is mainly saturated by two-gluon
glueballs. The fit gives
\begin{eqnarray}
\frac{T_c}{\sqrt{\sigma}}=0.71247
\,, \qquad {\rm Lattice} \quad 0.629(3)  \,,
\end{eqnarray}
a quite reasonable result. 

We can proceed further in the analysis by rewriting 
the trace anomaly for bosons as 
\begin{eqnarray}
\DeltaA(T) &=&
\sum_{k=1}^\infty \int dM \frac{\partial N(M)}{\partial M} \frac{1}{2 \pi^2 k} \left( \frac{M}{T} \right)^3 K_1\left(\frac{kM}{T}\right) \,, \nonumber 
\end{eqnarray} 
and making use of the formula 
\begin{eqnarray}
\int_0^\infty dM\, \left(\frac{M}{T} \right)^3 \frac{M^{n-1}}{2 \pi^2 k} K_1(k M /T)=
\frac{(n+2) (2 T)^n \Gamma \left(\frac{n}{2}+1\right)^2 
}{ 2 \pi^2 k^{n+4}}  \,,
\end{eqnarray}
as well as the large $M$ expansion of the cumulative number,
Eq.~(\ref{eq:largeM}). Then we get an equivalent large $T$ expansion
for the trace anomaly,
\begin{eqnarray}
\DeltaA(T)= \sum_{n} a_n 
\frac{  n (n+2) }
{2 \pi ^2} (2T)^n \zeta (n+4) \Gamma \left(\frac{n}{2}+1\right)^2 \, . 
\end{eqnarray}
For instance, in the two-gluon sector we get 
\begin{eqnarray}
\DeltaA_{2g}(T)=  \frac{2048 \pi^8}{3465} a_6 T^6 +
\frac{128 \pi^6}{1575} a_4 T^4 +  \frac{128 \pi^6}{1575} a_2 T^2  
.
\end{eqnarray}
Applying the results in Ref.~\cite{Caro:1994ht} for the coefficients $a_i$
yields
\begin{eqnarray}
\DeltaA(T)
=
\frac{32768 \, \pi ^7 T^6}{113669325 \, \sigma ^3}
+
\frac{512 \, \pi ^5 \alpha_s  T^4}{127575 \, \sigma ^2}
+
\frac{16}{945} \pi ^4 T^2 \left(\frac{2 \alpha_s^2}{\pi \sigma}
-\frac{16}{81 \pi \sigma }\right) \,.
\end{eqnarray}

The interesting feature is that the effect of the full quantized
spectrum can be very well described by the semiclassical expansion for
the temperatures available on the lattice. Thus, we can use instead
the semiclassical expansion to fit the parameters and sidestep the
diagonalization. 

As a further remark let us note that a direct attempt to fit the
polynomial formula fails since there are too few data points and the
curve is smooth. The correlations implicit in the 2g-Hamiltonian among
the different coefficients are however enough to guide the fit. The
last data point is intriguing and the most obvious candidate to fill
the gap is by looking at three-gluon glueballs. The three gluon
potential on the lattice has been analyzed~\cite{Cardoso:2008sb} and
it has been found that the triangle is preferred to the starfish
configuration. In a semirelativistic framework three gluon glueballs
have been addressed in Ref. \cite{Mathieu:2008pb}. Thus we can take
the partonic Hamiltonian result which sets a scale separation between
2g-WKB and 3g glueballs,
\begin{eqnarray}
\DeltaA_{3 g}(T) \sim e^{-M_{3g}/T} \ll \DeltaA_{2g}(T) 
.
\end{eqnarray}
Whether or not $3g$ saturate the missing contribution, or even
multigluonic states are needed remains to be seen.  On the other hand,
let us note that using the lowest glueballs on the lattice we
found~\cite{Megias:2009mp} that while the cumulative number could,
after come coarse graining, be described as exponentially growing with
$T_H= 2.8 T_c$, the effect on the trace anomaly was tiny. The
High-Precision Thermodynamics in connection to the Hagedorn Density of
States has been discussed in \cite{Meyer:2009tq}. A satisfactory fit
with $T_H=1.024(3)T_c$ was obtained by adding a string motivated
Hagedorn spectrum to the lowest $0^{++}$ and $2^{++}$ glueballs (see
also \cite{Buisseret:2011fq} and \cite{Borsanyi:2012ve} ). The fate of
glueballs above the phase transition has been analyzed in
Ref.~\cite{Lacroix:2012pt}.

\subsection{The adjoint Polyakov loop and the gluelump spectrum}

According to duality the adjoint Polyakov loop can be computed in the
confined phase in terms of the gluelump spectrum. In the simplest
case, this system corresponds to one massless spin-1 particle and one
gluon source (infinitely heavy) which is the CM system. The Salpeter
equation for the Hamiltonian operator (the Coulomb term is omitted) reads
\begin{eqnarray}
H \psi_n = (p  + \sigma_A r) \psi_n = M_n \psi_n  \, , 
\end{eqnarray}
which by rescaling the coordinate can be brought to the glueball spectrum, 
\begin{eqnarray}
M_{\rm gluelump} = M_{\rm glueball}/\sqrt{2} \, .
\end{eqnarray}
Thus, the smallest mass gap in the pure gauge theory is the gluelump
and not the glueball! This scaling relation of the spectrum
implies that the partition function fulfills
$ Z_{\rm gluelumps} (T) = (2/g)Z_{\rm glueballs} (T/\sqrt{2})$. Likewise,
for the adjoint Polyakov loop at low temperatures implies, 
\begin{eqnarray}
\langle \Omega_8 \rangle_T \sim Z_{\rm gluelumps} (T) = \sum_n e^{-\Delta_n/T} \neq 0 \qquad (T < T_c)  \,.
\end{eqnarray}

\subsection{The powers of deconfinement}

One straightforward application of the glueball gas is that because
they are heavy, the pressure in the confined phase is tiny, 
\begin{eqnarray}
P(T)=P_{\rm glueball}(T) \approx e^{-M_G/T}  \qquad (T<T_c) \,,
\end{eqnarray}
where $M_G \gg T_c$. On the other hand, at high temperatures the
pressure is due to a free gluon gas,
\begin{equation}
P_{\rm gluons}(T)= \frac{b_0}2 T^4 ,\qquad b_0 = \frac{
  2\pi^2}{45} (N_c^2-1) \qquad (T \gg T_c) \,.
\end{equation}
When the trace anomaly for gluodynamics was first evaluated on the
lattice~\cite{Boyd:1996bx} across the phase transition, the common
wisdom was the standard textbook explanation of the deconfinement
based on the MIT bag model, where hadrons are bubbles in the strongly
interacting vacuum with an energy density $B \sim M_H/V \sim 1 {\rm GeV}/{\rm fm}^3 \sim (0.3 {\rm GeV})^4$~\cite{Johnson:1975zp}. Thus
the total pressure was
\begin{eqnarray}
P(T)= P_{\rm gluons}(T)-B \,, \qquad T > T_c  \,,
\end{eqnarray}
and continuity of the pressure implies $P_{\rm gluons}(T_c) =B $ for
$M_G \gg T_c$, and thus yields the trace anomaly
\begin{eqnarray}
\DeltaA  = \frac{\epsilon-3P}{T^4} = \frac{4 B}{T^4}, \qquad T > T_c \,.
\end{eqnarray}
However, this behaviour is in complete disagreement with the
old~\cite{Boyd:1996bx} and newest~\cite{Borsanyi:2012ve} lattice data.

Ten years ago we looked into the Polyakov loop and found that,
contrary to expectations, one has inverse temperature power
corrections above the phase transition of the form of
Eq.~(\ref{eq:LT-dim2})~\cite{Megias:2005ve}, with a remarkable good
description of the data in a wide range of temperatures. These power
corrections were quite surprising and completely
unexpected,\footnote{In fact, a $1/T^2$ behaviour in the trace anomaly
  data of \cite{Boyd:1996bx} was already noted by the authors of
  \cite{Meisinger:2001cq}.} since it indicated the breakdown of
perturbation theory for temperatures as large as $5 T_c$, but have
been verified on the lattice~\cite{Gupta:2007ax,Mykkanen:2012ri} and
other models~\cite{Andreev:2009zk}.

One possibility to explain the trace anomaly data is to assume instead that
there is a temperature dependence in the bag constant (a fuzzy
bag~\cite{Pisarski:2006yk}, see
also~\cite{Megias:2008dv,Megias:2008rm,Megias:2009mp,Megias:2009ar} for a
unified setup).
\begin{equation}
P(T)= P_{\rm gluons}(T)-B_{\rm fuzzy} (T) ~~ (T > T_c ) , \quad 
P(T_c) = P_{\rm glueballs} (T_c)=0  .
\end{equation}
An inspiring consequence from the power corrections of Eq.~(\ref{eq:LT-dim2})
is to assume
\begin{eqnarray}
B_{\rm fuzzy} = \frac{b_0}2 T_c^2 T^2  \qquad  \longrightarrow \qquad P=
\frac{b_0}2 (T^4- T^2 T_c^2)
,
\end{eqnarray}
which gives
\begin{eqnarray}
\DeltaA = \frac{\epsilon-3P}{T^4}  = b_0 \left( \frac{T_c}{T} \right)^2 
, \qquad
b_0 = 3.51
\,.
\end{eqnarray} 

\begin{figure}
\epsfig{figure=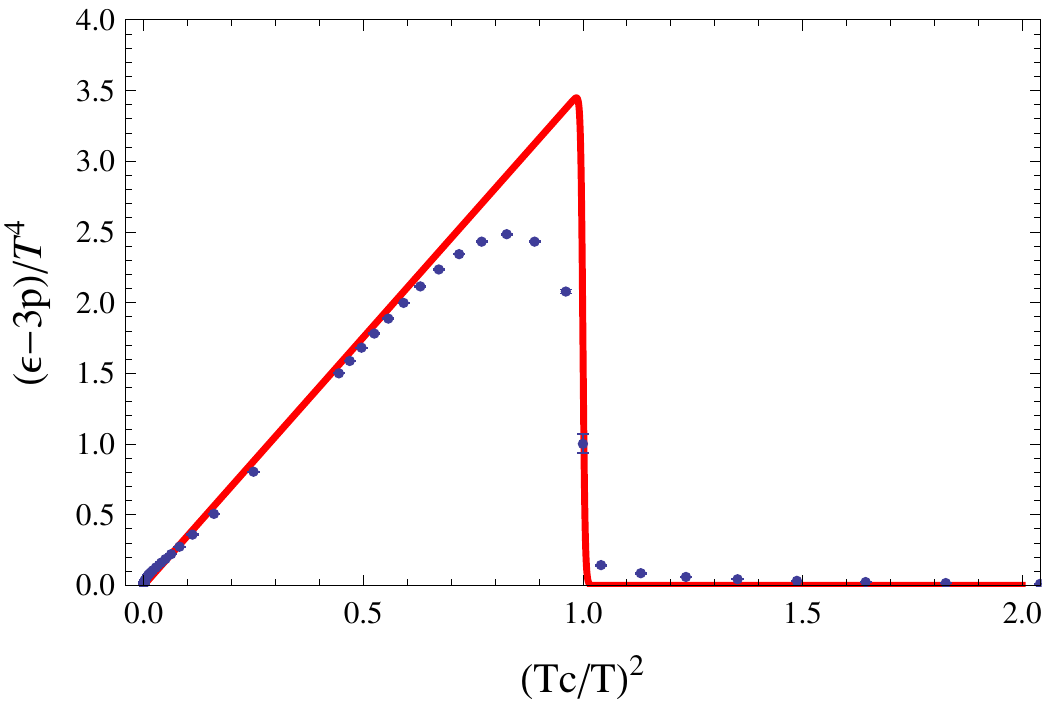,height=5cm,width=6cm,angle=0}
\epsfig{figure=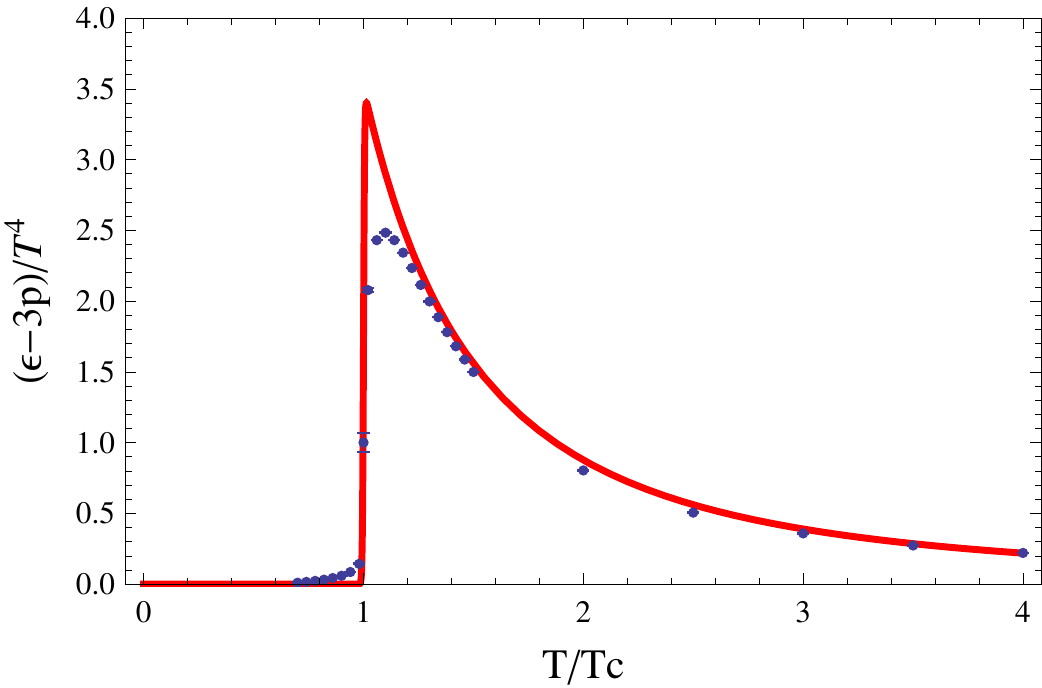,height=5cm,width=6cm,angle=0}
\caption{Trace anomaly as a function of $1/T^2$ and $T$ comparing the fuzzy
  bag of Pisarski~\cite{Pisarski:2006yk} $ \DeltaA(T) = \frac{(N_c^2-1)
    \pi^2}{45} \left( \frac{T_c}{T} \right)^2 \theta(T-T_c) $ with
  the lattice data from the WB collaboration~\cite{Borsanyi:2012ve}.}
\label{fig:fuzzy}
\end{figure}
The result is presented in Fig.~\ref{fig:fuzzy} and compared with the
most recent lattice data from the WB
collaboration~\cite{Borsanyi:2012ve}.  As we see the agreement for $T>
1.5 \, T_c$ of such a simple model is impressive. These features are
confirmed for any $N_c$~\cite{Panero:2009tv,Datta:2010sq}. Further
analyses involve renormalization group
resumations~\cite{Megias:2009mp}, hard thermal
loops~\cite{Andersen:2011ug} as well as holographic
methods~\cite{Gursoy:2009jd,Megias:2010ku,Zuo:2014iza}, but a clear
physical picture of the undoubtful but mysterious powers is still
lacking.

\section{Emergence of duality with Polyakov loop models}

\subsection{Chiral quark models at finite temperature}

Up to now, we have assumed that duality holds without specifying how
this might occur from a microscopic point of view. Here we want to
illustrate how duality arises from quark models and the relevant
provisos to achieve this goal. A full QCD argument for the
hadronization of the Polyakov loop was advanced and elaborated in
Refs. \cite{Megias:2012kb,Megias:2013xaa}.

Chiral quark models have been used (e.g.~\cite{Christov:1991se} and
references therein) for a long time to describe the chiral phase
transition. It was then realized that they lead to a wrong $N_c$
counting of thermal corrections, since they are described as one quark
loop and hence are ${\cal O}(N_c)$. In reality, they correspond to
zero point energy of meson states which are ${\cal O} (N_c^0)$. For
instance, in the quark condensate one has $\langle \bar q q \rangle_0
= {\cal O} (N_c)$ but $\langle \bar q q \rangle_T- \langle \bar q q
\rangle_0 = {\cal O} (N_c^0)$ (wrongly counted by the CQM as ${\cal O}
(N_c)$). This observation was well known although surprisingly nothing
was done about it. The requirement of large gauge invariance motivated
us the introduction of the Polyakov line as an independent
variable~\cite{Megias:2004hj} which has a local and quantum
character. Previous works had already dealt with this
coupling~\cite{Meisinger:1995ih,Fukushima:2003fw} initiating the
Polyakov--Nambu--Jona-Lasinio (PNJL)
saga~\cite{Ratti:2005jh,Sasaki:2006ww,Ciminale:2007sr,Contrera:2007wu,Schaefer:2007pw,Costa:2008dp,Mao:2009aq,Sakai:2010rp,Radzhabov:2010dd,Zhang:2010kn}
where the Polyakov line has been taken global and classical. This
allowed to study the interplay between chiral symmetry restoration and
breakdown of the center symmetry although, as we have repeatedly
pointed out in our previous
works~\cite{Megias:2004hj,Megias:2006bn,Megias:2005qf,Megias:2006df,Megias:2006ke},
this generates an undesirable ambiguity of group coordinates as well
as a non-vanishing value for the Polyakov loop in the
adjoint-representation, contradicting lattice
simulations.\footnote{See Refs.~\cite{Braun:2007bx,Braun:2009gm} for a
  study of this interplay from effective potential methods.} These
difficulties may be overcome~\cite{Megias:2004hj,Megias:2006bn} by
recognizing the local and quantum nature of the Polyakov loop.  Using
this interpretation we have shown that a gateway to the hadron
resonance gas may be established if the motion of quarks in the field
generated by the Polyakov loop is quantized.

We note that on the lattice~\cite{Langelage:2010yn} the HRG has been
deduced in strong coupling and for large $N_c$ by considering hadrons
at rest.

\subsection{The Quantum and local Polyakov loop}

As we have mentioned above, large gauge invariance is an important
restriction at finite temperature which breaks down in perturbation
theory. In the Polyakov gauge one can automatically implement large
gauge invariance by considering the Polyakov loop line $\Omega(\vec
x)$ as an independent variable, which in the Polyakov gauge becomes a
diagonal unitary matrix in colour space. This is equivalent to a
minimal coupling scheme in the time derivative of a dynamical quark or
gluon field. We refer to
Refs.~\cite{Megias:2004hj,Megias:2006bn,Megias:2005qf,Megias:2006df}
for further motivation.

Following this prescription, the partition function of the Chiral
Quark Model (CQM) at finite temperature can be written as 
\begin{equation}
Z_{\rm CQM}= \int D \Omega \, e^{-S(T, \Omega)} \,,
\label{eq:3}
\end{equation}
where $\Omega = e^{i g A_0/T}$ and $D \Omega$ is the invariant
$\SU(N_c)$ Haar group integration measure, for each $\SU(N_c)$ variable
$\Omega(\vec x)$ at each point $\vx$.  Here the action is
\begin{equation}
S (T,\Omega) =  S_q (T,\Omega) +  S_G (T,\Omega)  \,.
\end{equation}
The fermionic contribution depends on the quarks (and anti-quarks),
and it is obtained from the corresponding fermion
determinant. Assuming mass-degenerated quarks for simplicity the
effective action reads
\begin{eqnarray}
S_q (T,\Omega) =  -2 N_f \sum_q  \int \frac{d^3
  x d^3 p}{(2\pi)^3} \bigg( 
 \tr_c \log \big[ 1+\Omega ({\vec x}) \, e^{-E_{q}(p)/T}\big] +  {\rm c.c.}
\big] 
\bigg)  \,.
\label{eq:quark-action}
\end{eqnarray}
where c.c. stands for the complex conjugated contribution stemming
from the anti-quarks and 2 is the spin degeneracy factor for spin 1/2
particles.  Here $E_{q}(p) = \sqrt{\vp^2+M_q^2}$ is the energy of a
quark with total mass $M_q=M_0 +m_q$, and $M_0$ is the constituent
mass. As one can see the diagonal part of the Polyakov loop
corresponds to consider chemical potentials for different color
species. Large color gauge invariance is implemented by just averaging
over group elements. We will check whether this minimal coupling
scheme for the Polyakov line complies with known QCD properties
\cite{Megias:2012kb,Megias:2013xaa}, such as the fact that the
expectation value of the Polyakov loop admits a hadronic
representation.

\subsection{Hadronic representation of the Polyakov loop}

In order to see this, consider a system with $N_f$ dynamical quarks
and an extra heavy quark (not antiquark) of an arbitrarily large mass
$m_H$ at rest located at a fixed point and with fixed spin and colour
$a=1, \dots , N_c$.  From Eq.~(\ref{eq:quark-action}) the change in the
effective action is
\begin{eqnarray}
S_{q} (N_f+ 1) - S_q (N_f) &=& -2 \log (1+ \Omega_{aa} e^{-E_h/T} )  
\approx -2 e^{-m_H /T} \Omega_{aa}  \,,
\end{eqnarray}
yielding the partition function
\begin{eqnarray}
\frac{Z_{\rm CQM}^{\rm H}(N_f+ 1)}{Z_{\rm CQM}(N_f)} &=&  1+\langle \Omega_{aa} \rangle 2 e^{-m_H/T}  + \dots \nonumber \\ 
&=& 1+  \frac1{N_c}\langle \tr_c \Omega \rangle 2 e^{-m_H/T}   + \dots 
\end{eqnarray}
after averaging over color degrees of freedom implied by $D\Omega$. Thus we get 
\begin{eqnarray}
\frac1{N_c}\langle \tr_c \Omega \rangle = \lim_{m_H \to \infty} \frac12 \left[\frac{Z^{\rm H}_{\rm CQM}(N_f+ 1)}{Z_{\rm CQM}(N_f)} -1 \right] e^{m_H/T} \,.
\end{eqnarray}
To evaluate the r.h.s. we explicitly separate in the corresponding HRG
hadrons composed of hadrons made of $N_f$ dynamical quarks and one
extra heavy quark. The mass of such hadrons can be written as
\begin{eqnarray}
M_{q \dots , H}= \Delta_{q \dots } + m_H \,,
\end{eqnarray}
where by definition in $\Delta$ it has been subtracted the mass of the
heavy quark. On the other hand, the HRG partition function with
$N_f+1$-flavors with this extra heavy quark $H$ can be separated into
hadrons containing it or not. To do this, we reinstate the finite box
quantization conditions on the momentum $\vp$ to make sense of the
limit $m_H \to \infty$ and $V \to \infty$ (the Compton wavelength of
the heavy quark is shorter than the box size) and get
\begin{equation}
\log Z_{\rm HRG}^H(N_f+1) = \log Z_{\rm HRG}(N_f) +  \sum_{\vp,\alpha} \eta_\alpha  g_\alpha \log \left[1+ \eta_\alpha e^{-(\Delta_\alpha+m_H)/T} \right] \,.
\end{equation}
In the limit $m_H \to \infty$ only the states with $\vp=0$ survive
corresponding to a heavy Hadron at rest contribution, and thus we get
the result
\begin{eqnarray}
\frac12 \sum_\alpha g_\alpha e^{-\Delta_\alpha/T} = \lim_{m_H \to \infty} \frac12 \left[\frac{Z^{\rm H}_{\rm HRG}(N_f+ 1)}{Z_{\rm HRG}(N_f)} -1 \right] e^{m_H/T} \,.
\end{eqnarray}
Quark-Hadron duality at this level implies $Z_{\rm HRG} = Z_{\rm CQM}$
so that we get, 
\begin{eqnarray}
\frac1{N_c}\langle \tr_c \Omega \rangle = \frac12 \sum_\alpha g_\alpha e^{-\Delta_\alpha/T} \,,
\end{eqnarray}
providing confidence on the assumed minimal coupling of the Polyakov
line to quarks.

\subsection{From chiral quark models to the hadron resonance gas}

The previous models can be pictured as multiquark states which are
created or annihilated at point $\vec x$ and momentum $\vec p$ with
factors $ \Omega (\vec x) e^{-E_p/T} $ and $\Omega (\vec x)^\dagger
e^{-E_p/T}$. This corresponds to classical particles with internal
quantum numbers and statistic, i.e. to a second quantized but not
first quantized formalism.

At low temperatures quark Boltzmann factors are small $e^{-E_p/T} \ll 1$, and
the quark contribution to the action becomes small
\begin{eqnarray}
S_q [\Omega] =  -2 N_f  \int \frac{d^3 x d^3 p}{(2\pi)^3} 
\left[\tr_c \Omega (x) + \tr_c \Omega^\dagger (x) \right] e^{-E_p/T}+ \cdots \,.
\end{eqnarray}
Thus one has 
\begin{eqnarray}
Z_{\rm CQM} &=& \int D\Omega \, e^{-(S_q[\Omega]+S_G[\Omega])} = 
\langle e^{-S_q[\Omega]} \rangle_G \nonumber \\
&&\qquad\qquad=  \left\langle 1- S_q[\Omega] + \frac12 S_q[\Omega]^2 + \dots \right\rangle_G \,,
\end{eqnarray}
where $\langle \; \rangle_G $ incorporates besides the group
integration measure a gluon piece which needs not be specified at this
point. This expansion corresponds to a partonic expansion in terms of
constituents $q,\bar q , \bar q q , \dots$. The lowest non-vanishing
$\bar q q $ contribution reads
\begin{eqnarray}
Z_{\bar q q} &=&(2 N_f)^2  
\int \frac{d^3 x_1 d^3 p_1}{(2\pi)^3} 
\int \frac{d^3 x_2 d^3 p_2}{(2\pi)^3}  e^{-E_1/T} e^{-E_2/T}  \underbrace{\langle \tr_c \Omega (\vec x_1) \tr_c \Omega^\dagger (\vec x_2) \rangle_G}_{e^{-\sigma|\vec x_1-\vec x_2|/T}}   \nonumber \\ 
&=& (2 N_f)^2  \int 
\frac{d^3 x_1 d^3 p_1}{(2\pi)^3} 
\frac{d^3 x_2 d^3 p_2}{(2\pi)^3} e^{-H(x_1,p_1;x_2,p_2)/T}  \,,
\end{eqnarray}
where the $\bar q q $ Hamiltonian reads 
\begin{eqnarray}
H(x_1,p_1;x_2,p_2)= E_1 + E_2 + V_{12} \,.
\end{eqnarray}
Note that in the CM frame we get the same Salpeter equation we
discussed previously. Quantization in the CM frame $p_1 = - p_2 \equiv p$ leads to
\begin{eqnarray}
\left( 2 \sqrt{p^2 + M^2} + V_{q \bar q} (r) \right) \psi_n = M_n \psi_n \,,
\end{eqnarray}
and Boosting the CM to any frame with momentum $P$ we get the result 
\begin{eqnarray}
Z_{\bar q q} \to 
\sum_n \int \frac{d^3 R d^3 P}{(2\pi)^3} e^{-\frac{\sqrt{M_n^2+P^2}}{T}} \,,
\end{eqnarray}
which corresponds to the lowest order in a gas of non-interacting
mesons~\cite{RuizArriola:2012wd}. We have checked that this
equivalence holds up to $\bar q q \bar q q $ contributions, but fails
for higher Fock state components~\cite{Megias:2013xaa}. The reason has
to do with an ambiguity on what states should be considered colour
irreducible, i.e.  those in which all constituents are needed to
screen the source, without additional constituents forming a color
singlet by themselves. Our analysis faces, once more, the difficulties
in making a clear cut definition of a hadronic state out of
multiparton states.

The Polyakov loop can be treated in a similar way, 
\begin{eqnarray}
\frac1{N_c}\langle \tr_c \Omega \rangle =  &=& 2 N_f
\int \frac{d^3 x \, d^3 p}{(2\pi)^3} 
e^{-E_p/T}
\frac{1}{N_c} \underbrace{\langle \tr_c \Omega (\vec x_0) \tr_c \Omega^\dagger (\vec x) \rangle_G}_{e^{-\sigma|\vec x_0-\vec x|/T}}  
+\cdots
\nonumber \\ 
&=& \frac{2 N_f}{N_c} \int \frac{d^3 x \, d^3 p }{(2\pi)^3} e^{-H(\vec x,\vp)/T} \to\frac{2 N_f}{N_c}  \sum_n e^{-\Delta_n/T} \,,
\end{eqnarray}
where we have quantized the Heavy-light ground state system as
\begin{equation}
H(p, x) \psi_n= \left( \sqrt{p^2+m_q^2} + \sigma r \right) \psi_n = \Delta_n \psi_n  \,.
\end{equation} 
In previous works the quantization of the quark motion was not
considered, and as a consequence we failed to see the connection to
the HRG. In particular, we found that $L_T \sim e^{-M_0/T}$ so we
proposed to determine the constituent quark mass from an analysis of
the Polyakov loop on the lattice at low temperatures.\footnote{As we
  have noted above, in the heavy quark limit one has $L_T \sim
  e^{-m_Q/T}$.}  While fits along these lines turned out to provide
too large a constituent mass we preferred to keep the $L_T \sim
e^{-M_0/T}$ behaviour with a suitable proportionality
factor~\cite{Megias:2005qf}. As we have seen here the Boltzmann factor
contains the gap $\Delta$ (the heavy-light meson mass) which according
to our discussion above corresponds to {\it twice} the constituent
quark mass, so that $L_T \sim e^{- 2 M_0/T}$ and now the value for
$M_0$ from the lattice is phenomenologically acceptable.

A similar argument holds for the correlation function between Polyakov
loops. This requires an assumption for a four-point correlator in the
pure gluonic theory which at low temperature we assume to be
\begin{eqnarray}
\langle \tr_c \Omega (\vec x)^\dagger \tr_c \Omega (0) \tr_c \Omega
(\vec x_1)^ \dagger \tr_c \Omega (\vec x_2) \rangle_G &=& e^{-\sigma
  r/T} e^{-\sigma r_{12}/T} \nonumber \\ &+& e^{-\sigma |\vec x- \vec x_2|/T} e^{-\sigma
  r_1/T}
\end{eqnarray}
and satisfies cluster decomposition properties. This yields a
disconnected piece due to the effect of the quark determinant,
\begin{eqnarray}
\langle \tr_c \Omega (\vec x ) \tr_c \Omega^\dagger (0) \rangle &=& \frac{e^{-\sigma r /T} \left[ 1 + Z_{\bar q q} + \dots \right]+ | \langle \tr_c \Omega \rangle |^2}{1+Z_{\bar q q} + \dots} \nonumber \\ 
&=& e^{-\sigma r /T} + | \langle \tr_c \Omega \rangle |^2 \, . 
\end{eqnarray}
This reproduces the free energy formula Eq.~(\ref{eq:pol-corr-f})
based on a avoided crossing structure of the $\bar Q Q$ energy levels,
see Fig.~\ref{fig:Avoided-X}, and yields the result for $F_1(r,T)$
sketched in Fig.~\ref{fig:F1}.

\subsection{Gluon models with Polyakov loop}

We may introduce gluon fields besides those involved by the Polyakov
line variable. A particularly interesting case is gluodynamics.  At
one loop the effective potential has been computed
in~\cite{Weiss:1980rj}. More recently, one loop gluon actions with
Polyakov loops in the adjoint representation were suggested in
Refs.~\cite{Meisinger:2003id,Ruggieri:2012ny,Sasaki:2012bi,Sasaki:2013hsa},
where the background gauge for the classical gluon-field was assumed
and a Polyakov gauge for the quantum fluctuations in the field.  We
consider this form of dynamics but keeping our interpretation of a
quantum and local Polyakov loop variable. The partition function is
\begin{eqnarray}
Z= \int D \Omega w[\Omega] Z[\Omega] \equiv \langle Z[\Omega] \rangle \,,
\end{eqnarray}
where we assume the two-point correlation function to be
\begin{eqnarray}
\langle {\rm tr}_A \Omega (\vec x){\rm tr}_A \Omega (\vec y) \rangle =
e^{-\sigma_A |\vec x -\vec y|/T} .
\end{eqnarray}
According to our previous discussion and foreseeing the quantization
of partonic degrees of freedom, we write in compact form the action as
follows (see also \cite{Megias:2013xaa}),
\begin{eqnarray}
\log Z[ \Omega]=-{\rm Tr} \log \left( 1- e^{-h/T}\right) \,,
\label{eq:logZglue}
\end{eqnarray}
where ${\rm Tr} = \int d^3 x \, \tr_A \sum_{\lambda=\pm}$
and the single particle Hamiltonian is given by 
\begin{eqnarray}
h = p - i g A_0(\vec x) \,, \qquad  \Omega(\vec x)= e^{i g A_0(\vec x)/T} \,.
\end{eqnarray}
The interpretation of the previous formula is that of a particle in a
random purely imaginary gluon field with given correlation functions,
and it has been written in a way that preserves large gauge invariance
in the Polyakov gauge.  In the semiclassical approximation we can
replace the quantum mechanical trace as we already did
before,\footnote{Note that we have scalar gluons with 2 spin
  states. This is {\it not} 3 spin-1 states nor 2 helicity states.}
\begin{eqnarray}
\log Z [\Omega] = -2 \int \frac{d^3 x d^3 p}{(2\pi)^3} {\rm tr}_A \log \left( 1- e^{-p/T} \Omega (\vec x) \right) \,,
\end{eqnarray}
where the factor 2 comes from the two spin states. In the limit of
a classical and global $\Omega$ we get the action of
Refs.~\cite{Meisinger:2003id,Ruggieri:2012ny,Sasaki:2012bi,Sasaki:2013hsa}.
The similar criticisms on the ambiguities on the choice of group
coordinates at the mean field level apply here. The compact notation
in Eq.~(\ref{eq:logZglue}) is extremely useful to carry out the
partonic expansion analysis at low temperatures and pursue the mapping
to the glueball gas. Thus we have to compute
\begin{eqnarray}
Z= \left\langle \exp \left[ {\rm Tr} \log \left( 1- e^{-h/T}\right) \right] \right\rangle 
\end{eqnarray}
in a power expansion of $e^{-h/T}$.  At lowest order we have $ \langle
{\rm tr} \, e^{-h/T} \rangle =0 $, and the next order yields,
\begin{eqnarray}
\frac12 \langle \left( {\rm tr} \, e^{-h/T}\right)^2 \rangle &=&  
\frac{g^2}2 \int \frac{d^3 x_1 d^3 p_1}{(2\pi)^3} \frac{d^3 x_2 d^3 p_2}{(2\pi)^3}
e^{-H(p_1,x_1; p_2,x_2)/T} \nonumber \\ 
&\to& {\rm Tr}_2 \, e^{-H_2/T} \,,
\end{eqnarray}
where $H_2$ is the two-gluon Hamiltonian whose spectrum provides the
two gluon glueballs discussed in the previous section, and ${\rm
  Tr}_2$ is the trace in the corresponding two-gluon Hilbert
space. Note that the factor 1/2 corresponds to the correct Boltzmann
counting which originates in the quantum statistical mechanics to
avoid the Gibbs paradox in the high-temperature
limit~\cite{Huang:1987bk}. Following similar steps as before we can
obtain the gluelump representation of the Polyakov loop in the adjoint
representation.  One can analyze what happens for higher Fock states
since the glueball gas corresponds to having
\begin{eqnarray}
\log Z = - \sum_n {\rm Tr}_n \log \left( 1- e^{-H_n/T}\right)  \,,
\end{eqnarray}
where here ${\rm Tr}= \sum_n {\rm Tr}_n$ represents the trace over the
whole multigluon Hilbert space. Pursuing the low temperature expansion
to higher orders we have checked that up to three gluons included, the
mapping with the glueball mass works but ambiguities arise for 4-gluon
states, where the possibility of forming separate and weakly
interacting 2-gluon glueballs first arises. We are facing again the
concept of colour irreducible clusters inside colour neutral
states~\cite{Megias:2013xaa} and the very definition of a hadron.

\section{Conclusions}

Quark-Hadron Duality at finite temperature is the statement that at
low temperatures hadrons can be considered a complete basis of
states. The naive hadron resonance gas, while simple minded, works
well enough at sufficiently high temperatures as to deserve dedicated
attention on {\it why} becomes this picture invalid. This success
remains a mystery over the years since Hagedorn first proposed it.

The listed PDG states incorporate currently just the $q \bar q$ or
$qqq$ states which fit into the conventional quark model, but what is
the nature of states that are needed when approaching the crossover
from below? As we have seen, saturating at subcritical temperatures
requires many hadronic states, and so the excited spectrum involves
relativistic effects even for heavy quarks. This point makes
relativistic quark models a potential source for investigation of the
hadron spectrum from a global and thermodynamic perspective since the
number of needed excited states challenges any lattice QCD
calculation.  Moreover, the hadronic mapping of Polyakov loops in
fundamental and higher $\SU(N_c)$ colour group representations allows
to deduce multiquark states, gluelumps and hybrid states, containing
one or several heavy quark or gluon sources. This goes beyond the
models and opens up the possibility of a Polyakov loop spectroscopy
including exotics.  While the question on {\it what} is the complete
hadronic particle spectrum remains still open, we envisage the
possibility of grasping the yet unknown physics of the phase
transition by enquiring this question at the lowest possible
temperature where the hadron resonance gas picture fails.

%\begin{acknowledgments}
\bigskip
\bigskip
One of us (E.R.A.) warmly thanks Michal Prasza\l owicz for the
invitation and for providing a stimulating atmosphere. We also thank
Wojtek Broniowski for many discussions and Marco Panero for
clarifications. This work has been supported by Plan Nacional de Altas
Energ\'{\i}as FPA2011-25948, DGI FIS2011-24149, Junta de
Andaluc\'{\i}a grant FQM-225, Generalitat de Catalunya grant
2014-SGR-1450, Spanish MINECO's Consolider-Ingenio 2010 Programme CPAN
(CSD2007-00042) and Centro de Excelencia Severo Ochoa Programme grant
SEV-2012-0234. The research of E.M. has been supported by the Juan de
la Cierva Program of the Spanish MINECO, and by the European Union
under a Marie Curie Intra-European Fellowship (FP7-PEOPLE-2013-IEF).
%\end{acknowledgments}
\bigskip

%\bibliographystyle{h-elsevier}
%\bibliography{Refs,zakopane}

\begin{thebibliography}{100}

\bibitem{Kogut:1982rt}
J.B. Kogut et~al.,
\newblock Phys.Rev.Lett. 50 (1983) 393.

\bibitem{Polonyi:1984zt}
J. Polonyi et~al.,
\newblock Phys.Rev.Lett. 53 (1984) 644.

\bibitem{Pisarski:1983ms}
R.D. Pisarski and F. Wilczek,
\newblock Phys.Rev. D29 (1984) 338.

\bibitem{Aoki:2006we}
Y. Aoki et~al.,
\newblock Nature 443 (2006) 675, hep-lat/0611014.

\bibitem{Philipsen:2012nu}
O. Philipsen,
\newblock Prog.Part.Nucl.Phys. 70 (2013) 55, 1207.5999.

\bibitem{Florkowski:2010zz}
W. Florkowski,
\newblock {Phenomenology of Ultra-Relativistic Heavy-Ion Collisions} (World
  Scientific, Singapore, 2010).

\bibitem{Borsanyi:2013bia}
S. Borsanyi et~al.,
\newblock Phys.Lett. B730 (2014) 99, 1309.5258.

\bibitem{Bazavov:2014pvz}
HotQCD Collaboration, A. Bazavov et~al.,
\newblock (2014), 1407.6387.

\bibitem{Melnitchouk:2005zr}
W. Melnitchouk, R. Ent and C. Keppel,
\newblock Phys.Rept. 406 (2005) 127, hep-ph/0501217.

\bibitem{Lucini:2012gg}
B. Lucini and M. Panero,
\newblock Phys.Rept. 526 (2013) 93, 1210.4997.

\bibitem{Gerber:1988tt}
P. Gerber and H. Leutwyler,
\newblock Nucl.Phys. B321 (1989) 387.

\bibitem{Dashen:1969ep}
R. Dashen, S.K. Ma and H.J. Bernstein,
\newblock Phys.Rev. 187 (1969) 345.

\bibitem{Venugopalan:1992hy}
R. Venugopalan and M. Prakash,
\newblock Nucl.Phys. A546 (1992) 718.

\bibitem{Kostyuk:2000nx}
A. Kostyuk et~al.,
\newblock Phys.Rev. C63 (2001) 044901, hep-ph/0004163.

\bibitem{Nakamura:2010zzi}
Particle Data Group, K. Nakamura et~al.,
\newblock J. Phys. G37 (2010) 075021.

\bibitem{Godfrey:1985xj}
S. Godfrey and N. Isgur,
\newblock Phys.Rev. D32 (1985) 189.

\bibitem{Capstick:1986bm}
S. Capstick and N. Isgur,
\newblock Phys.Rev. D34 (1986) 2809.

\bibitem{Bazavov:2013dta}
A. Bazavov et~al.,
\newblock Phys. Rev. Lett. 111, 082301 (2013) 082301, 1304.7220.

\bibitem{Huovinen:2009yb}
P. Huovinen and P. Petreczky,
\newblock Nucl.Phys. A837 (2010) 26, 0912.2541.

\bibitem{Sekihara:2012xp}
T. Sekihara and T. Hyodo,
\newblock Phys.Rev. C87 (2013) 045202, 1209.0577.

\bibitem{Begun:2012rf}
V. Begun, M. Gazdzicki and M. Gorenstein,
\newblock Phys.Rev. C88 (2013) 024902, 1208.4107.

\bibitem{Arriola:2012vk}
E.Ruiz~Arriola, W. Broniowski and P. Masjuan,
\newblock Acta Phys.Polon.Supp. 6 (2013) 95, 1210.7153.

\bibitem{Suzuki:1994ay}
T. Suzuki et~al.,
\newblock Phys.Lett. B347 (1995) 375, hep-lat/9408003.

\bibitem{Dunne:1996yb}
G.V. Dunne, K.M. Lee and C.h. Lu,
\newblock Phys.Rev.Lett. 78 (1997) 3434, hep-th/9612194.

\bibitem{Salcedo:1998sv}
L. Salcedo,
\newblock Nucl.Phys. B549 (1999) 98, hep-th/9802071.

\bibitem{Boyd:1996bx}
G. Boyd et~al.,
\newblock Nucl.Phys. B469 (1996) 419, hep-lat/9602007.

\bibitem{Leutwyler:1992cd}
H. Leutwyler,
\newblock (1992).

\bibitem{Miller:2006hr}
D.E. Miller,
\newblock Phys.Rept. 443 (2007) 55, hep-ph/0608234.

\bibitem{Svetitsky:1985ye}
B. Svetitsky,
\newblock Phys.Rept. 132 (1986) 1.

\bibitem{Kaczmarek:2002mc}
O. Kaczmarek et~al.,
\newblock Phys.Lett. B543 (2002) 41, hep-lat/0207002.

\bibitem{Megias:2005ve}
E. Megias, E. Ruiz~Arriola and L. Salcedo,
\newblock JHEP 0601 (2006) 073, hep-ph/0505215.

\bibitem{Gava:1981qd}
E. Gava and R. Jengo,
\newblock Phys.Lett. B105 (1981) 285.

\bibitem{Kaczmarek:2005uv}
O. Kaczmarek and F. Zantow,
\newblock Eur.Phys.J. C43 (2005) 63, hep-lat/0502011.

\bibitem{Megias:2012kb}
E. Megias, E. Ruiz~Arriola and L. Salcedo,
\newblock Phys.Rev.Lett. 109 (2012) 151601, 1204.2424.

\bibitem{Gupta:2007ax}
S. Gupta, K. Huebner and O. Kaczmarek,
\newblock Phys.Rev. D77 (2008) 034503, 0711.2251.

\bibitem{Mykkanen:2012ri}
A. Mykkanen, M. Panero and K. Rummukainen,
\newblock JHEP 1205 (2012) 069, 1202.2762.

\bibitem{Megias:2013xaa}
E. Megias, E. Ruiz~Arriola and L. Salcedo,
\newblock Phys.Rev. D89 (2014) 076006, 1311.2814.

\bibitem{Megias:2014bfa}
E. Megias, E.Ruiz~Arriola and L. Salcedo,
\newblock (2014), 1409.0773.

\bibitem{Kapusta:1979fh}
J.I. Kapusta,
\newblock Nucl.Phys. B148 (1979) 461.

\bibitem{Borsanyi:2012ve}
S. Borsanyi et~al.,
\newblock JHEP 1207 (2012) 056, 1204.6184.

\bibitem{Greensite:2003bk}
J. Greensite,
\newblock Prog.Part.Nucl.Phys. 51 (2003) 1, hep-lat/0301023.

\bibitem{Luscher:2002qv}
M. Luscher and P. Weisz,
\newblock JHEP 0207 (2002) 049, hep-lat/0207003.

\bibitem{West:1995ym}
G.B. West,
\newblock Phys.Rev.Lett. 77 (1996) 2622, hep-ph/9603316.

\bibitem{Coester:1965zz}
F. Coester,
\newblock Helv.Phys.Acta 38 (1965) 7.

\bibitem{Keister:1991sb}
B. Keister and W. Polyzou,
\newblock Adv.Nucl.Phys. 20 (1991) 225.

\bibitem{Keister:2011ie}
B. Keister and W. Polyzou,
\newblock Phys.Rev. C86 (2012) 014002, 1109.6575.

\bibitem{hakim2011introduction}
R. Hakim,
\newblock Introduction to Relativistic Statistical Mechanics: Classical and
  Quantum (World Scientific, 2011).

\bibitem{Anzai:2010td}
C. Anzai, Y. Kiyo and Y. Sumino,
\newblock Nucl.Phys. B838 (2010) 28, 1004.1562.

\bibitem{Bali:2000un}
G.S. Bali,
\newblock Phys.Rev. D62 (2000) 114503, hep-lat/0006022.

\bibitem{Bali:2005fu}
SESAM Collaboration, G.S. Bali et~al.,
\newblock Phys.Rev. D71 (2005) 114513, hep-lat/0505012.

\bibitem{landau1965quantum}
L.D. Landau and E.M. Lifshits,
\newblock Quantum Mechanics Non-relativistic Theory: Transl. from the Russian
  by JB Sykes and JS bell. 2d Ed., rev. and Enl (Pergamon Press, 1965).

\bibitem{Dashen:1974ns}
R.F. Dashen and G.L. Kane,
\newblock Phys.Rev. D11 (1975) 136.

\bibitem{Kaczmarek:2005ui}
O. Kaczmarek and F. Zantow,
\newblock Phys.Rev. D71 (2005) 114510, hep-lat/0503017.

\bibitem{Arriola:2013jxa}
E. Ruiz~Arriola, L. Salcedo and E. Megias,
\newblock Acta Phys.Polon.Supp. 6 (2013) 953, 1304.2245.

\bibitem{Bazavov:2011nk}
A. Bazavov et~al.,
\newblock Phys. Rev. D85 (2012) 054503, 1111.1710.

\bibitem{Borsanyi:2010bp}
Wuppertal-Budapest Collaboration, S. Borsanyi et~al.,
\newblock JHEP 1009 (2010) 073, 1005.3508.

\bibitem{Hagedorn:1965st}
R. Hagedorn,
\newblock Nuovo Cim.Suppl. 3 (1965) 147.

\bibitem{Broniowski:2000bj}
W. Broniowski and W. Florkowski,
\newblock Phys.Lett. B490 (2000) 223, hep-ph/0004104.

\bibitem{Broniowski:2004yh}
W. Broniowski, W. Florkowski and L.Y. Glozman,
\newblock Phys.Rev. D70 (2004) 117503, hep-ph/0407290.

\bibitem{Cohen:2011cr}
T.D. Cohen and V. Krejcirik,
\newblock J.Phys. G39 (2012) 055001, 1107.2130.

\bibitem{Johnson:1975zp}
K. Johnson,
\newblock Acta Phys.Polon. B6 (1975) 865.

\bibitem{Kapusta:1981ay}
J.I. Kapusta,
\newblock Phys.Rev. D23 (1981) 2444.

\bibitem{Arriola:2011en}
E. Ruiz~Arriola and W. Broniowski,
\newblock (2011) 7, 1110.2863.

\bibitem{Masjuan:2012gc}
P. Masjuan, E. Ruiz~Arriola and W. Broniowski,
\newblock Phys.Rev. D85 (2012) 094006, 1203.4782.

\bibitem{Masjuan:2013xta}
P. Masjuan, E. Ruiz~Arriola and W. Broniowski,
\newblock Phys.Rev. D87 (2013) 118502, 1305.3493.

\bibitem{Masjuan:2012sk}
P. Masjuan, E. Ruiz~Arriola and W. Broniowski,
\newblock Phys.Rev. D87 (2013) 014005, 1210.0760.

\bibitem{Karl:1999wq}
G. Karl and J.E. Paton,
\newblock Phys.Rev. D60 (1999) 034015, hep-ph/9904407.

\bibitem{Simonov:2000ky}
Y. Simonov,
\newblock Nucl.Phys. B592 (2001) 350, hep-ph/0003114.

\bibitem{Guo:2007sm}
P. Guo et~al.,
\newblock Phys.Rev. D77 (2008) 056005, 0707.3156.

\bibitem{Marsh:2013xsa}
K. Marsh and R. Lewis,
\newblock Phys.Rev. D89 (2014) 014502, 1309.1627.

\bibitem{Meyer:2004gx}
H.B. Meyer,
\newblock (2004), hep-lat/0508002.

\bibitem{Gregory:2012hu}
E. Gregory et~al.,
\newblock JHEP 1210 (2012) 170, 1208.1858.

\bibitem{Mathieu:2008me}
V. Mathieu, N. Kochelev and V. Vento,
\newblock Int.J.Mod.Phys. E18 (2009) 1, 0810.4453.

\bibitem{Jacob:1959at}
M. Jacob and G. Wick,
\newblock Annals Phys. 7 (1959) 404.

\bibitem{Barnes:1981ac}
T. Barnes,
\newblock Z.Phys. C10 (1981) 275.

\bibitem{Mathieu:2008bf}
V. Mathieu, F. Buisseret and C. Semay,
\newblock Phys.Rev. D77 (2008) 114022, 0802.0088.

\bibitem{Buisseret:2009ea}
F. Buisseret,
\newblock Phys.Rev. D79 (2009) 037503, 0902.1028.

\bibitem{Mathieu:2009cc}
V. Mathieu,
\newblock PoS QCD-TNT09 (2009) 024, 0910.4855.

\bibitem{Morningstar:1999rf}
C.J. Morningstar and M.J. Peardon,
\newblock Phys.Rev. D60 (1999) 034509, hep-lat/9901004.

\bibitem{Szczepaniak:1995cw}
A. Szczepaniak et~al.,
\newblock Phys.Rev.Lett. 76 (1996) 2011, hep-ph/9511422.

\bibitem{Szczepaniak:2003mr}
A.P. Szczepaniak and E.S. Swanson,
\newblock Phys.Lett. B577 (2003) 61, hep-ph/0308268.

\bibitem{Meyers:2012ka}
J. Meyers and E.S. Swanson,
\newblock Phys.Rev. D87 (2013) 036009, 1211.4648.

\bibitem{Swanson:2013hta}
E. Swanson,
\newblock Acta Phys.Polon.Supp. 6 (2013) 859.

\bibitem{Bicudo:2007wt}
P. Bicudo,
\newblock Phys.Rev. D76 (2007) 094005, hep-ph/0703114.

\bibitem{Caro:1994ht}
J. Caro, E. Ruiz~Arriola and L. Salcedo,
\newblock J.Phys. G22 (1996) 981, nucl-th/9410025.

\bibitem{Cardoso:2008sb}
M. Cardoso and P. Bicudo,
\newblock Phys.Rev. D78 (2008) 074508, 0807.1621.

\bibitem{Mathieu:2008pb}
V. Mathieu, C. Semay and B. Silvestre-Brac,
\newblock Phys.Rev. D77 (2008) 094009, 0803.0815.

\bibitem{Megias:2009mp}
E. Megias, E. Ruiz~Arriola and L. Salcedo,
\newblock Phys.Rev. D80 (2009) 056005, 0903.1060.

\bibitem{Meyer:2009tq}
H.B. Meyer,
\newblock Phys.Rev. D80 (2009) 051502, 0905.4229.

\bibitem{Buisseret:2011fq}
F. Buisseret and G. Lacroix,
\newblock Phys.Lett. B705 (2011) 405, 1105.1092.

\bibitem{Lacroix:2012pt}
G. Lacroix et~al.,
\newblock Phys.Rev. D87 (2013) 054025, 1210.1716.

\bibitem{Meisinger:2001cq}
P.N. Meisinger, T.R. Miller and M.C. Ogilvie,
\newblock Phys.Rev. D65 (2002) 034009, hep-ph/0108009.

\bibitem{Andreev:2009zk}
O. Andreev,
\newblock Phys.Rev.Lett. 102 (2009) 212001, 0903.4375.

\bibitem{Pisarski:2006yk}
R.D. Pisarski,
\newblock Prog.Theor.Phys.Suppl. 168 (2007) 276, hep-ph/0612191.

\bibitem{Megias:2008dv}
E. Megias, E. Ruiz~Arriola and L. Salcedo,
\newblock Indian J.Phys. 85 (2011) 1191, 0805.4579.

\bibitem{Megias:2008rm}
E. Megias, E. Ruiz~Arriola and L. Salcedo,
\newblock Nucl.Phys.Proc.Suppl. 186 (2009) 256, 0809.2044.

\bibitem{Megias:2009ar}
E. Megias, E. Ruiz~Arriola and L. Salcedo,
\newblock Phys.Rev. D81 (2010) 096009, 0912.0499.

\bibitem{Panero:2009tv}
M. Panero,
\newblock Phys.Rev.Lett. 103 (2009) 232001, 0907.3719.

\bibitem{Datta:2010sq}
S. Datta and S. Gupta,
\newblock Phys.Rev. D82 (2010) 114505, 1006.0938.

\bibitem{Andersen:2011ug}
J.O. Andersen et~al.,
\newblock Phys.Rev. D84 (2011) 087703, 1106.0514.

\bibitem{Gursoy:2009jd}
U. Gursoy et~al.,
\newblock Nucl.Phys. B820 (2009) 148, 0903.2859.

\bibitem{Megias:2010ku}
E. Megias, H. Pirner and K. Veschgini,
\newblock Phys.Rev. D83 (2011) 056003, 1009.2953.

\bibitem{Zuo:2014iza}
F. Zuo,
\newblock JHEP 1406 (2014) 143, 1404.4512.

\bibitem{Christov:1991se}
C. Christov, E. Ruiz~Arriola and K. Goeke,
\newblock Acta Phys.Polon. B22 (1991) 187.

\bibitem{Megias:2004hj}
E. Megias, E. Ruiz~Arriola and L. Salcedo,
\newblock Phys.Rev. D74 (2006) 065005, hep-ph/0412308.

\bibitem{Meisinger:1995ih}
P.N. Meisinger and M.C. Ogilvie,
\newblock Phys.Lett. B379 (1996) 163, hep-lat/9512011.

\bibitem{Fukushima:2003fw}
K. Fukushima,
\newblock Phys.Lett. B591 (2004) 277, hep-ph/0310121.

\bibitem{Ratti:2005jh}
C. Ratti, M.A. Thaler and W. Weise,
\newblock Phys.Rev. D73 (2006) 014019, hep-ph/0506234.

\bibitem{Sasaki:2006ww}
C. Sasaki, B. Friman and K. Redlich,
\newblock Phys. Rev. D75 (2007) 074013, hep-ph/0611147.

\bibitem{Ciminale:2007sr}
M. Ciminale et~al.,
\newblock Phys. Rev. D77 (2008) 054023, 0711.3397.

\bibitem{Contrera:2007wu}
G.A. Contrera, D. Gomez~Dumm and N.N. Scoccola,
\newblock Phys. Lett. B661 (2008) 113, 0711.0139.

\bibitem{Schaefer:2007pw}
B.J. Schaefer, J.M. Pawlowski and J. Wambach,
\newblock Phys. Rev. D76 (2007) 074023, 0704.3234.

\bibitem{Costa:2008dp}
P. Costa et~al.,
\newblock Phys. Rev. D79 (2009) 116003, 0807.2134.

\bibitem{Mao:2009aq}
H. Mao, J. Jin and M. Huang,
\newblock J. Phys. G37 (2010) 035001, 0906.1324.

\bibitem{Sakai:2010rp}
Y. Sakai et~al.,
\newblock Phys. Rev. D82 (2010) 076003, 1006.3648.

\bibitem{Radzhabov:2010dd}
A. Radzhabov et~al.,
\newblock Phys.Rev. D83 (2011) 116004, 1012.0664.

\bibitem{Zhang:2010kn}
T. Zhang, T. Brauner and D.H. Rischke,
\newblock JHEP 06 (2010) 064, 1005.2928.

\bibitem{Megias:2006bn}
E. Megias, E. Ruiz~Arriola and L. Salcedo,
\newblock Phys.Rev. D74 (2006) 114014, hep-ph/0607338.

\bibitem{Megias:2005qf}
E. Megias, E. Ruiz~Arriola and L. Salcedo,
\newblock PoS JHW2005 (2006) 025, hep-ph/0511353.

\bibitem{Megias:2006df}
E. Megias, E. Ruiz~Arriola and L. Salcedo,
\newblock AIP Conf.Proc. 892 (2007) 444, hep-ph/0610095.

\bibitem{Megias:2006ke}
E. Megias, E. Ruiz~Arriola and L. Salcedo,
\newblock Eur.Phys.J. A31 (2007) 553, hep-ph/0610163.

\bibitem{Braun:2007bx}
J. Braun, H. Gies and J.M. Pawlowski,
\newblock Phys.Lett. B684 (2010) 262, 0708.2413.

\bibitem{Braun:2009gm}
J. Braun et~al.,
\newblock Phys.Rev.Lett. 106 (2011) 022002, 0908.0008.

\bibitem{Langelage:2010yn}
J. Langelage and O. Philipsen,
\newblock JHEP 1004 (2010) 055, 1002.1507.

\bibitem{RuizArriola:2012wd}
E. Ruiz~Arriola, E. Megias and L. Salcedo,
\newblock AIP Conf.Proc. 1520 (2013) 185, 1207.4875.

\bibitem{Weiss:1980rj}
N. Weiss,
\newblock Phys.Rev. D24 (1981) 475.

\bibitem{Meisinger:2003id}
P.N. Meisinger, M.C. Ogilvie and T.R. Miller,
\newblock Phys.Lett. B585 (2004) 149, hep-ph/0312272.

\bibitem{Ruggieri:2012ny}
M. Ruggieri et~al.,
\newblock Phys.Rev. D86 (2012) 054007, 1204.5995.

\bibitem{Sasaki:2012bi}
C. Sasaki and K. Redlich,
\newblock Phys.Rev. D86 (2012) 014007, 1204.4330.

\bibitem{Sasaki:2013hsa}
C. Sasaki,
\newblock Acta Phys.Polon.Supp. 7 (2014) 107, 1312.3818.

\bibitem{Huang:1987bk}
K. Huang,
\newblock Statistical Mechanics (Addison Wesley \& Sons, New York, 1987).

\end{thebibliography}

\end{document}